\documentclass[aps,prd,twocolumn,showpacs,superscriptaddress,nofootinbib,preprintnumbers]{revtex4-2}

\usepackage{amssymb,amsmath,verbatim,mathtools,needspace,enumitem,etoolbox,graphicx,physics,microtype,afterpage,xspace,tabularx,lmodern,multirow}
\usepackage{booktabs}
\usepackage{siunitx}
\usepackage{gensymb}
\usepackage{ulem}  

\usepackage{amsmath}
\usepackage{tikz}
\usepackage{tikz-feynman}
\usepackage{booktabs}
\usepackage{bm}
\usepackage{cancel}
\tikzfeynmanset{
  compat=1.1.0
}
\tikzset{every path/.append style={thick}}
\usepackage{amsfonts}
\usepackage{amssymb}
\usepackage{enumerate}
\usepackage{color}
\usepackage{graphicx}
\usepackage{bm}
\usepackage[colorlinks=true,citecolor=blue,urlcolor=blue]{hyperref}
\hypersetup{colorlinks=true,citecolor=romared,linkcolor=romared,urlcolor=romared}
\definecolor{romared}{RGB}{142,0,28}
\usepackage{afterpage}
\usepackage{multirow}
\usepackage{appendix}
\usepackage[table]{xcolor}
\usepackage{feynmp-auto}
\usepackage{aas_macros}
\usepackage{makecell}
\usepackage{soul}

\usepackage{lipsum}
\usepackage{orcidlink}
\definecolor{tabblue}{RGB}{31, 119, 180}
\definecolor{darkblue}{RGB}{0, 0, 120}
\definecolor{tabred}{RGB}{214, 39, 40}
\definecolor{tabgreen}{RGB}{44, 160, 44}
\definecolor{tabgray}{RGB}{100, 100, 100}
\definecolor{taborange}{RGB}{255, 127, 14}
\definecolor{tabbrown}{RGB}{128, 0, 0}
\definecolor{tabpink}{RGB}{255, 141, 161}
\definecolor{tabpurple}{RGB}{148, 103, 189}
\definecolor{goldenrod}{RGB}{218, 165, 32}
\usepackage{float}
\usepackage{placeins}
\usepackage{verbatim}
\newcommand{\Neff}{N_\mathrm{eff}}

\begin{document}
\preprint{\hbox{UTWI-23-2025}}

\title{Directly Probing Neutrino Interactions through CMB Phase Shift Measurements}
\author{Gabriele Montefalcone\,\orcidlink{0000-0002-6794-9064}}
\email{montefalcone@utexas.edu}
\affiliation{Texas Center for Cosmology and Astroparticle Physics, Weinberg Institute for Theoretical Physics, Department of Physics, The University of Texas at Austin, Austin, TX 78712, USA}

\author{Subhajit Ghosh\,\orcidlink{0000-0002-6973-4673}}
\email{sghosh@utexas.edu}
\affiliation{Texas Center for Cosmology and Astroparticle Physics, Weinberg Institute for Theoretical Physics, Department of Physics, The University of Texas at Austin, Austin, TX 78712, USA}

\author{Kimberly K.~Boddy\,\orcidlink{0000-0003-1928-4667}}
\affiliation{Texas Center for Cosmology and Astroparticle Physics, Weinberg Institute for Theoretical Physics, Department of Physics, The University of Texas at Austin, Austin, TX 78712, USA}

\author{Daven Wei Ren Ho\,\orcidlink{0009-0006-6376-0383
}}
\affiliation{Department of Physics and Astronomy, University of Notre Dame, IN 46556, USA}

\author{Yuhsin Tsai\,\orcidlink{0000-0001-7847-225X}}
\affiliation{Department of Physics and Astronomy, University of Notre Dame, IN 46556, USA}
 
\begin{abstract}
\noindent Perturbations in the cosmic neutrino background produce a characteristic phase shift in the acoustic oscillations imprinted in the anisotropies of the cosmic microwave background (CMB), providing a unique observational probe of neutrino physics. In this work, we explore how this phase shift signature is altered in the presence of neutrino interactions with temperature-dependent scattering rates, motivated by physical constructions for neutrino self-interactions and neutrino-dark matter couplings. A key finding is that the phase shift in these realistic models---characterized by gradual rather than instantaneous decoupling---maintains the same functional form as the free-streaming template, with only the asymptotic amplitude decreasing for stronger interactions that delay decoupling. This simple parametrization enables us to directly constrain neutrino interactions through phase shift measurements in the temperature and polarization power spectra from CMB observations. Analyzing the latest data from \textit{Planck}, the Atacama Cosmology Telescope, and the South Pole Telescope, we derive strong constraints on the neutrino decoupling redshift. Our global analysis indicates that neutrinos have been freely streaming since deep within the radiation-dominated epoch. We also explore flavor-dependent scenarios in which only one neutrino species interacts. Overall, our work establishes a signature-driven framework that exploits the clean phase shift signal in the acoustic oscillations of the CMB as a precise and robust probe of non-standard neutrino interactions in the early universe.
\end{abstract}
 
\maketitle
\tableofcontents
\section{Introduction}

\noindent The anisotropies of the cosmic microwave background (CMB) provide a direct window into the physics of the early universe and are particularly sensitive to the energy density and physical properties of the primordial radiation bath. A crucial component of this radiation bath is neutrinos. In the standard cosmological model, neutrinos decouple from the primordial plasma at temperatures around $1\,{\rm MeV}$, roughly one second after the Big Bang, and have been freely streaming through the cosmos ever since, forming the cosmic neutrino background~(C$\nu$B; see e.g.\ Ref.~\cite{Scott:2024rwc} for a recent review).

The energy density of the C$\nu$B is commonly parametrized through the effective number of relativistic neutrino species
\begin{equation}
    N_{\rm eff}=\frac{8}{7}\left(\frac{11}{4}\right)^{4/3}\frac{\rho_\nu}{\rho_\gamma}\equiv a_\nu \frac{\rho_\nu}{\rho_\gamma}\;,
    \label{eq:Neff}
\end{equation}
where $\rho_\gamma$ and $\rho_\nu$ are the photon and neutrino energy density, respectively, and $a_\nu\approx 4.40$. In the Standard Model (SM), this parameter takes the value $N_{\rm eff}^{\rm SM} = 3.044$~\cite{Akita:2020szl, Froustey:2020mcq, Bennett:2020zkv, Cielo:2023bqp, Drewes:2024wbw, Drewes:2024nbg}, accounting for three neutrino species with small corrections arising primarily from the careful consideration of neutrino decoupling. While direct detection of the C$\nu$B remains experimentally challenging~\cite{Bauer:2022lri, PTOLEMY:2019hkd}, neutrino properties can be inferred from their gravitational imprint on cosmological observables.

In particular, due to their free-streaming nature, neutrinos induce a characteristic phase shift in the acoustic oscillations of the CMB~\cite{Bashinsky:2003tk,Follin:2015hya,Baumann:2015rya}. Neutrino perturbations propagate at nearly the speed of light after they decouple from the primordial plasma, much faster than the corresponding sound waves in the tightly coupled photon-baryon fluid. The gravitational pull from these fast-moving neutrino perturbations in turn shifts the
photon and baryon perturbations toward slightly larger scales.  The resulting phase shift manifests as a multipole-dependent displacement of the acoustic peaks in the CMB temperature and polarization power spectra, with a specific scale dependence that is uniquely difficult to mimic through modifications to initial conditions or other cosmological parameters~\cite{Baumann:2015rya}. In addition, assuming adiabatic initial conditions, such a coherent shift in the CMB power spectra can only be produced by free-streaming radiation~\cite{Baumann:2015rya}, making the phase shift an exceptionally robust and clean signature of the free-streaming nature of neutrinos.

This distinctive signature was first detected directly in the \textit{Planck} 2013 temperature power spectrum~\cite{Follin:2015hya} and indirectly in \textit{Planck} 2015 temperature and polarization power spectra~\cite{Baumann:2015rya}. Most recently, Ref.~\cite{Montefalcone:2025unv} detected a nonzero phase shift at $14\sigma$ significance and found consistency with the SM prediction of three free-streaming neutrinos at the $1\sigma$ level. These precise measurements already indirectly constrain non-standard neutrino interactions beyond the weak force, as any sizable self-interactions or couplings to the dark sector would alter the free-streaming behavior of neutrinos. In particular, strong interactions in the neutrino sector would decrease the propagation speed of neutrino perturbations after weak decoupling, producing non-trivial modifications to both the amplitude and multipole dependence of the induced phase shift in the CMB acoustic oscillations.

In this paper, we derive the phase-shift templates in realistic neutrino interaction models and develop an analysis pipeline that directly constrains the redshift of neutrino decoupling from phase shift measurements in CMB power spectra. A key finding of our work is that interactions characterized by gradual decoupling produce phase shift signatures that can still be captured within a remarkably simple framework.
Specifically, we show that the phase shift is well-described by the same functional form as in the free-streaming case, with only its asymptotic amplitude rescaled according to the neutrino decoupling redshift. The later the decoupling occurs, the smaller the induced phase shift becomes, with the amplitude smoothly interpolating between the free-streaming and fluid-like limits.

The precise mapping between this asymptotic amplitude and the decoupling redshift depends on the temperature scaling of the interaction rate. Here, we focus on two classes of interactions with scattering rates $\Gamma_\nu\propto T_\nu^3$ and $\Gamma_\nu\propto T_\nu^5$, where $T_\nu$ is the neutrino bath temperature. The $T_\nu^3$ scaling arises in models of neutrino-dark matter (DM) interactions~\cite{Boehm:2003hm,Stadler:2019dii,Serra:2009uu,Olivares-DelCampo:2017feq,Ghosh:2017jdy}, while the $T_\nu^5$ scaling characterizes both neutrino self-interactions~\cite{Cyr-Racine:2013jua,Oldengott:2014qra,Oldengott:2017fhy,Kreisch:2019yzn,Barenboim:2019tux,Das:2020xke,Das:2023npl,Loverde:2022wih,RoyChoudhury:2020dmd,Lancaster:2017ksf,Camarena:2023cku,Poudou:2025qcx,He:2023oke,He:2025jwp} and neutrino-DM coupling scenarios~\cite{Wilkinson:2014ksa,Escudero:2015yka,DiValentino:2017oaw,Ghosh:2019tab,Mosbech:2020ahp,He:2025jwp}. The straightforward parametrization of the phase-shift template in terms of decoupling redshift enables us to extend the direct phase shift measurement framework~\cite{Follin:2015hya,Montefalcone:2025unv} to probe these neutrino interaction models via a robust signature-driven approach. Importantly, the different temperature scalings lead to distinct mappings between the asymptotic amplitude and decoupling redshift, providing a potential avenue to distinguish between interaction scenarios.

Using the latest temperature and polarization datasets from the \textit{Planck} satellite, the Atacama Cosmology Telescope (ACT), and the South Pole Telescope (SPT), we place strong constraints on the neutrino decoupling redshift using phase-shift measurements in both flavor-universal scenarios and flavor-dependent interaction scenarios, where only a fraction of neutrino species participate in the interactions.

Throughout this work, the effects of neutrino self-interactions are computed with \texttt{nuCLASS}~\cite{Libanore:2025ack}\footnote{\href{https://github.com/subhajitghosh-phy/nuCLASS.git}{\texttt{https://github.com/subhajitghosh-phy/nuCLASS.git}}}, a modified version of the Boltzmann solver \texttt{CLASS}~\cite{Blas:2011rf}, while neutrino–DM scattering is modeled with \texttt{CLASS}'s built-in interacting DM–dark radiation module~\cite{Archidiacono:2019wdp}, based on the ETHOS framework~\cite{Cyr-Racine:2015ihg}. Appendix~\ref{sec:A1} provides further details on these implementations. In Appendices~\ref{sec:A2} through~\ref{sec:A4}, we include supplementary material covering the generalized phase-shift extraction from CMB power spectra with interacting neutrinos and validation of our template approximation, a comparison of our direct phase-shift analysis pipeline with established methods~\cite{Follin:2015hya,Montefalcone:2025unv}, and a complete summary of constraints from all CMB data combinations not shown in the main text.

Finally, we assume massless neutrinos. The massless limit is an excellent approximation in the early universe, given current cosmological bounds on neutrino masses, which have been shown to have a negligible impact on direct phase‑shift measurements from observed CMB power spectra~\cite{Montefalcone:2025unv}. The constraints derived in this work suggest that neutrino decoupling occurs in the ultra-relativistic regime, which further supports the massless neutrino assumption.

\begin{figure*}[ht!]
    \centering
    \includegraphics[width=.9\linewidth]{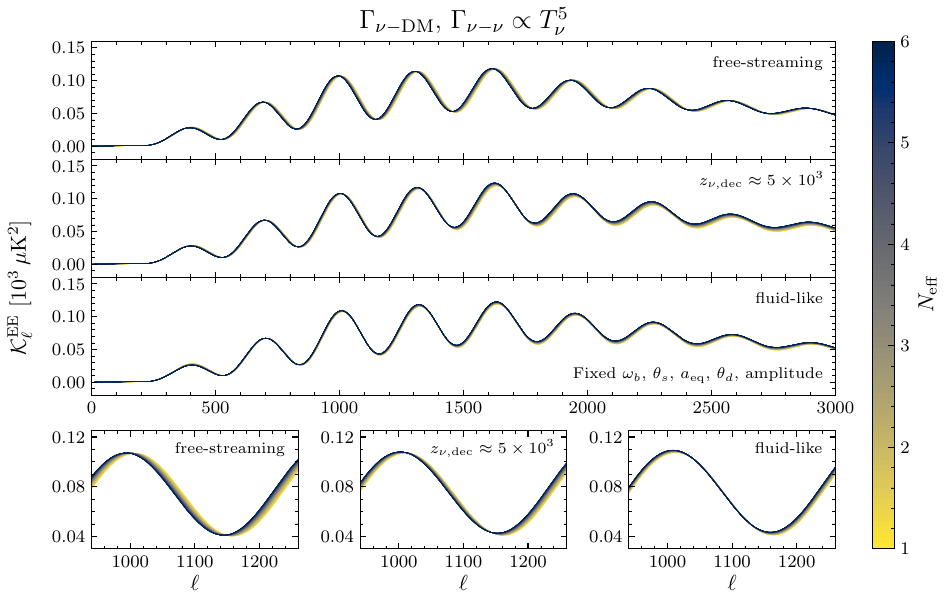}
    \caption{Illustration of the neutrino phase shift imprinted in the undamped polarization power spectrum $\mathcal{K}^{EE}_\ell$, Eq.~\eqref{eq:K_ell}, for interacting neutrinos with a scattering rate $\Gamma_\nu\propto T_\nu^5$. We show three distinct neutrino decoupling scenarios: (i) neutrinos effectively always free-streaming, corresponding to $z_{\nu,\rm dec}\gg10^{5}$; (ii) neutrinos decoupling close to recombination, specifically with $z_{\nu,\rm dec}\approx 5\times 10^{3}$; and (iii) fluid-like neutrinos throughout, corresponding to $z_{\nu,\rm dec}\ll10^{3}$. To clearly isolate the neutrino-induced phase shift parameterized by $\Neff$, we fix $\omega_b$, $a_{\mathrm{eq}}$, $\theta_s$, $\theta_d$ (specifically to our fiducial $\Lambda \rm CDM$ parameter values shown in Table~\ref{tab:parameters}), and the height of the fourth EE peak. The upper panels display the polarization power spectrum across the multipole range~$\ell$ relevant for current CMB experiments, while the lower panels zoom in around the fourth EE peak, highlighting the sensitivity of the phase shift to the neutrino decoupling epoch. An equivalent phase shift signature is also present in the TT and TE power spectra.}
    \label{fig:1}
\end{figure*}

\section{Phase Shift Signatures of Neutrino Interactions}\label{sec:2}

\noindent This section examines how the phase shift signature in the acoustic oscillations of the CMB evolves from the well-understood case of SM free-streaming neutrinos to the more complex scenarios involving interacting neutrinos that decouple at later cosmic epochs. We first review the theoretical framework for the neutrino-induced phase shift in the standard model of cosmology, where neutrinos have been freely streaming since redshift $z \sim 10^{10}$, before exploring how neutrino interactions alter this signature.

\subsection{Free-streaming Standard Model neutrinos}

\noindent Following their decoupling from the primordial plasma at redshift $z_{\rm SM\nu,\rm{dec}}\sim 10^{10}$, SM neutrinos free-stream at approximately the speed of light, which is greater than the sound speed $c_s\approx c/\sqrt{3}$ of the photon-baryon fluid.
Neutrino perturbations, therefore, induce metric fluctuations ahead of the sound horizon, effectively pulling the photon and baryon perturbations toward larger scales and causing the acoustic oscillations to acquire a phase shift $\phi$~\cite{Bashinsky:2003tk,Baumann:2015rya}.  This phase shift ultimately manifests as a multipole shift in the CMB power spectra that can be parametrized as~\cite{Follin:2015hya, Montefalcone:2025unv}
\begin{equation}
    \delta\ell(N_{\rm eff})= A(N_{\rm eff}) f_{\rm\ell},\label{eq:delta_ell}
\end{equation}
with
\begin{align}
    A(\Neff) &\equiv \frac{\epsilon(\Neff)-\epsilon(3.044)}{\epsilon(1)-\epsilon(3.044)}\, ,	\label{eq:ANeff} \\
    f_\ell &= \frac{\ell_{\rm \infty}}{1+(\ell/\ell_\star)^\xi},\label{eq:fell}
\end{align}
where $\epsilon(N_{\rm eff})\equiv \rho_\nu/\rho_r=N_{\rm eff}/\left(a_\nu +N_{\rm eff}\right)$ is the fractional energy density in neutrinos, $\rho_r=\rho_\gamma+\rho_\nu$ is the total radiation density.

The amplitude $A(\Neff)$ and template function $f_\ell$ characterize the overall size of the effect and its multipole dependence, respectively. $A(N_{\rm eff})$ is linear in the fractional energy density of neutrinos~\cite{Bashinsky:2003tk, Baumann:2015rya} and is normalized here in a manner consistent with previous studies~\cite{Follin:2015hya, Montefalcone:2025unv}.

The functional form of the template $f_\ell$ reflects the physical timescales governing the impact of neutrinos on the acoustic oscillations of the primordial plasma. Namely, it asymptotically approaches a constant $\ell_\infty$ at high multipoles, as expected for modes that enter the horizon deep in radiation domination, producing a constant phase shift.  Conversely, at low multipoles corresponding to modes that enter the horizon during matter domination, the shift smoothly approaches zero, governed by the parameters $\ell_\star$ and $\xi<0$. The vanishing phase shift reflects the diminishing influence of neutrinos as radiation becomes increasingly subdominant relative to matter~\cite{Baumann:2015rya}. For the case of SM neutrinos, the fitting function well-approximates the induced shifts in the acoustic peaks of the CMB power spectra for $\ell_\infty = 11.0 \pm 0.6$, $\ell_\star = 483 \pm 53$, and $\xi = -1.69 \pm 0.13$~\cite{Montefalcone:2025unv}.

\begin{figure*}
    \centering
    \includegraphics[width=\linewidth]{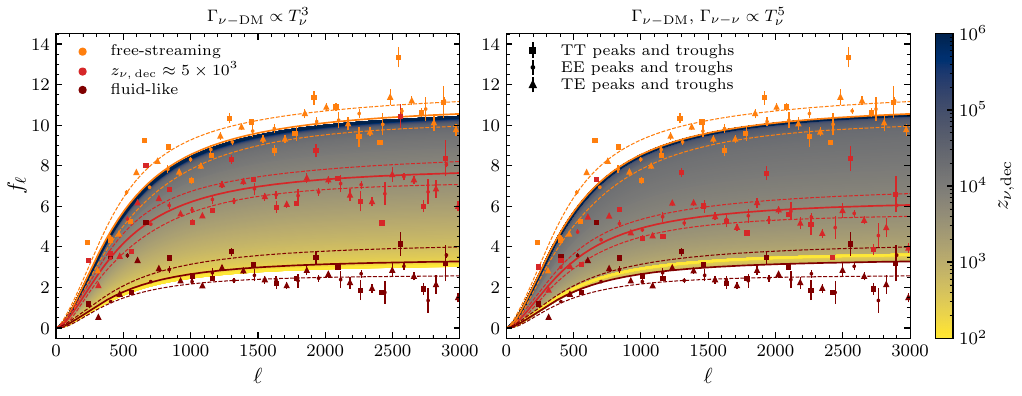}
    \caption{Spectrum-based templates $f_\ell$, as defined in Eqs.~\eqref{eq:fell} and \eqref{eq:fell_int}, as a function of multipole $\ell$ for interacting neutrinos with scattering rates $\Gamma_\nu \propto T_\nu^3$ ({\it left}) and $\Gamma_\nu \propto T_\nu^5$ ({\it right}). Solid \textcolor{taborange}{orange}, \textcolor{tabred}{red}, and \textcolor{tabbrown}{brown} lines indicate the best-fit templates for three representative scenarios: fully free-streaming neutrinos ($z_{\nu,\rm dec} \gg 10^5$), delayed decoupling ($z_{\nu,\rm dec} \approx 5\times 10^3$), and fluid-like behavior ($z_{\nu,\rm dec} \ll 10^3$). The corresponding $2\sigma$ confidence intervals are shown as dashed lines. For these benchmark scenarios, we also display the numerically extracted phase shifts for the TT (squares), EE (circles), and TE (triangles) power spectra, obtained by sampling 100 cosmologies with $N_{\rm eff} \in [1, 6]$ and normalized as described in Appendix~\ref{sec:A2}. The extracted phase shifts are color-coded to match the respective best-fit templates.}
    \label{fig:2}
\end{figure*}

\subsection{Interacting Neutrinos}\label{sec:2.2}
\noindent Thus far, we have assumed that neutrinos have always been freely streaming throughout the entire cosmic history. In the context of the SM, this approximation is robust, since neutrinos decouple from the primordial plasma at such a high redshift ($z_{\rm SM \nu,dec}\sim 10^{10}$) that they remain effectively collisionless throughout all redshifts relevant for CMB observations.

However, this picture is altered if neutrinos are strongly interacting, for instance, due to self-interactions or direct couplings to DM. The main effect of such interactions is to delay the time of decoupling, which in turn can significantly alter both the size and multipole-dependence of the neutrino-induced phase shift in the CMB.

Figure~\ref{fig:1} demonstrates the effect of the neutrino phase shift imprinted in the undamped polarization (EE) power spectrum, as defined in Eq.~\eqref{eq:K_ell}; an equivalent phase shift signature is also present in the temperature (TT) and cross-correlation (TE) power spectra. We show the case of neutrinos interactions with a scattering rate $\Gamma_\nu\propto T_\nu^5$, which is representative of known physical constructions for neutrino self-interactions and DM-$\nu$ scatterings~\cite{Boehm:2003hm,Wilkinson:2014ksa,Cyr-Racine:2013jua}, for different neutrino decoupling redshifts $z_{\nu,\rm dec}$.
To clearly isolate the neutrino-induced phase shift as a function of their energy density as parameterized by $\Neff$, we fix the physical baryon density~$\omega_b$, the scale factor at matter-radiation equality~$a_{\mathrm{eq}}$, the angular size of the sound horizon~$\theta_s$, the angular size of the damping scale~$\theta_d$, and the height of the fourth EE peak, following Refs.~\cite{Follin:2015hya, Wallisch:2018rzj, Montefalcone:2025unv}.

\subsubsection{Instantaneous decoupling approximation}

\noindent To build intuition, let us first consider the simplified case of instantaneous decoupling, where neutrinos transition abruptly from a tightly coupled fluid to being completely free-streaming at a redshift $z_{\nu,\rm dec}$, set by the strength of the new interaction. In this scenario, the phase shift follows the same template as Eq.~\eqref{eq:delta_ell} for all multipoles $\ell < \ell_{\rm dec}$, where $\ell_{\rm dec}$ corresponds to the scale relevant for modes entering the horizon at $z_{\rm dec}$.

However, for $\ell \gtrsim \ell_{\rm dec}$, the phase shift is significantly reduced because neutrinos behave as a fluid and propagate at their sound speed $c_{s,\nu}\approx c/\sqrt{3}$ rather than the speed of light~(see e.g. Ref.~\cite{Choi:2018gho} in the context of dark radiation decoupling).
Nevertheless, a nonzero phase shift persists at high multipoles as long as the neutrino sound speed exceeds that of the photon-baryon fluid. This condition is always satisfied for self-interacting (SI) neutrinos and remains valid when neutrinos couple to a small fraction of DM. 
Namely, for SI neutrinos $c_{s,\rm SI\nu}=c/\sqrt{3}>c_{s,\gamma}\equiv c/\sqrt{3\left(1+R_{b,\gamma}\right)}$ where $R_{b,\gamma}\equiv 3\rho_b/(4\rho_\gamma)$ and $\rho_b$ is the baryon energy density.

For neutrinos strongly coupled to a fraction $f_{\rm iDM}$ of DM, the propagation speed of the neutrino fluid can be slowed down further, an effect known as DM loading~\cite{Ghosh:2024wva}. Large interacting fractions $f_{\rm iDM} > 10\%$ can cause the neutrino sound speed to be smaller than that of the photon-baryon fluid, which induces a phase shift in the opposite direction to that of free-streaming neutrinos.
While a minor DM loading effect exists even for $f_{\rm iDM} < 10\%$ when $c_{s,\gamma}<c_{s,\rm iDM-\nu}<c/\sqrt{3}$, we set $f_{\rm iDM} = 10^{-3}$ throughout this paper to effectively neglect this contribution and focus purely on the modification of neutrino free-streaming properties, ensuring a direct comparison with the SI neutrino scenario~\cite{Ghosh:2017jdy,Ghosh:2019tab,Ghosh:2024wva}.

\subsubsection{Gradual decoupling from realistic neutrino interactions}\label{sec:3.2}

\noindent While the simplified picture of instantaneous decoupling provides useful intuition, its applicability is limited. For many neutrino interaction models, we find that the same intuition does not hold.  The neutrino scattering rate $\Gamma_\nu$ is generically proportional to a power law in $T_\nu$~\cite{Boehm:2003hm,Stadler:2019dii}, making the transition from coupled to uncoupled behavior gradual rather than abrupt. Around the decoupling redshift $z_{\nu,\rm dec}$, defined by the condition 
$\left[\Gamma_\nu/H\right](z_{\nu,\rm dec})=1$, the mean free-path of neutrinos remains comparable to the Hubble radius ($H^{-1}$), rather than effectively jumping from zero to values much larger than $H^{-1}$, as in the instantaneous decoupling approximation. Therefore, during this extended transition period when neutrinos are \textit{loosely-coupled}, their propagation speed takes intermediate values between the fluid sound speed, $c/\sqrt{3}$, and the free-streaming speed of light, $c$. The duration of this regime is determined by the specific temperature scaling of the interaction rate. Therefore, to accurately determine the final imprint on the CMB power spectra and to extract phase shift, we follow the full evolution of the perturbations.

Specifically, we compute the induced phase shift $\delta\ell$ directly from the lensed power spectra using \texttt{CLASS} and~\texttt{nuCLASS} (see Appendix~\ref{sec:A1} for implementation details) for cases in which $\Gamma_\nu\propto T^3_\nu$ and $\Gamma_\nu\propto T^5_\nu$. In a straightforward phenomenological model for SI neutrinos, interactions are mediated by a heavy mediator, yielding $\Gamma_{\nu-\nu} \propto T_\nu^5$~\cite{Cyr-Racine:2013jua,Oldengott:2014qra,Oldengott:2017fhy,Kreisch:2019yzn,Das:2020xke,Das:2023npl,Loverde:2022wih,RoyChoudhury:2020dmd,Lancaster:2017ksf,Camarena:2023cku,Poudou:2025qcx,He:2023oke,He:2025jwp}. Neutrino-DM scattering models produce both $\Gamma_{\nu-\rm DM} \propto T_\nu^3$ and
$\Gamma_{\nu-\rm DM} \propto T_\nu^5$, depending on whether the mediator mass is comparable to or much larger than the DM mass, respectively~\cite{Boehm:2003hm,Stadler:2019dii,Serra:2009uu,Olivares-DelCampo:2017feq,Ghosh:2017jdy,Ghosh:2024wva,Wilkinson:2014ksa,Escudero:2015yka,DiValentino:2017oaw,Ghosh:2019tab,Mosbech:2020ahp}.

\begin{figure}
    \centering
    \includegraphics[width=\linewidth]{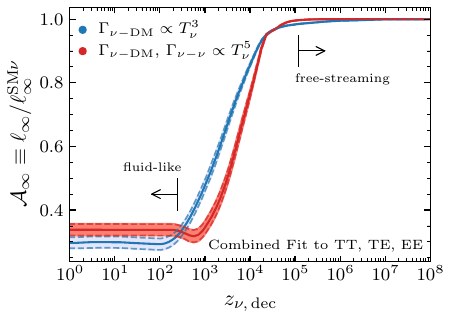}
    \caption{The amplitude ratio $\mathcal{A}_\infty$, defined as the asymptotic amplitude of the neutrino-induced phase-shift template $\ell_\infty$ relative to that of SM free-streaming neutrinos $\ell^{\rm SM \nu}_\infty$, shown as a function of neutrino decoupling redshift $z_{\nu,\rm dec}$. Results are obtained from the spectrum-based template fits $f_\ell$ using the lensed TT, TE, and EE power spectra. \textcolor{tabred}{Red} and \textcolor{tabblue}{blue} curves correspond to neutrino scattering rates scaling as $\Gamma_\nu \propto T_\nu^3$ and $\Gamma_\nu \propto T_\nu^5$, respectively. Solid lines indicate the corresponding best-fit values, while shaded regions depict their $2\sigma$ confidence intervals, further outlined by dashed lines. As expected, the template amplitude approaches the SM free-streaming limit, $\mathcal{A}_\infty = 1$, at high decoupling redshifts and the fluid-like limit, $\mathcal{A}_\infty \approx 0.3$, at low decoupling redshifts. The transition between these regimes is sharper in the $\Gamma_\nu \propto T_\nu^5$ scenario compared to the $\Gamma_\nu \propto T_\nu^3$ scenario, yet still spans 2-3 decades in redshift.}
    \label{fig:3}
\end{figure}

We find that in all of these scenarios, the induced phase shift in the CMB power spectra is well-approximated by the same functional form as Eq.~\eqref{eq:delta_ell}, with only the asymptotic value $\ell_\infty$ of the template function $f_\ell$ modified relative to the SM case:
\begin{equation}
    f_\ell=f_\ell^{\rm SM\nu}\times \mathcal{A}_\infty,  \label{eq:fell_int}  
\end{equation}
where $\mathcal{A_\infty}\equiv \ell_\infty/\ell^{\rm SM\nu}_\infty$ quantifies the amplitude modification relative to the SM scenario of free-streaming neutrinos. 
The resulting spectrum-based templates $f_\ell$ and the corresponding  amplitude ratios $\mathcal{A}_\infty$, are displayed in Figs.~\ref{fig:2} and~\ref{fig:3}, respectively, for both the $\Gamma_\nu\propto T^3_\nu$ and $\Gamma_\nu\propto T_\nu^5$ models as a function of decoupling redshift, $z_{\rm \nu, dec}$. To obtain these templates, we closely follow the prescription in Ref.~\cite{Montefalcone:2025unv} with minor modifications, detailed in Appendix~\ref{sec:A2}.

Figure~\ref{fig:2} illustrates explicitly how the instantaneous decoupling approximation substantially underestimates the induced phase shift in realistic models of neutrino interactions. We do not observe a sudden amplitude drop in the neutrino-induced phase shift for multipoles $\ell > \ell_{\rm dec}$, expected from instantaneous decoupling. Instead, the phase shift remains approximately constant at large multipoles, with the asymptotic value decreasing smoothly as the decoupling redshift decreases. The template amplitude asymptotes to the SM free-streaming limit, $\mathcal{A}_\infty =1$, at large $z_{\nu,\rm dec}$, and approaches the fluid-like regime with $\mathcal{A}_\infty \approx 0.3$ at small $z_{\nu,\rm dec}$.

The functional form of the template can be understood in terms of the propagation speed of neutrino perturbations during the {\it extended decoupling transition}. Prior to decoupling, when $\Gamma_\nu \gtrsim H$, the standard tightly-coupled fluid approximation is inadequate, as neutrino perturbations propagate at a much higher speed than the fluid sound speed, inducing a sizable phase shift in the acoustic oscillations of the photon-baryon fluid. Similarly, after decoupling when $\Gamma_\nu \lesssim H$, neutrinos remain under the influence of the interaction, which prevents them from fully free-streaming, leaving an extended window where their propagation speed stays below the speed of light.

Consequently, for modes entering the horizon deep in radiation domination, the induced phase shift by interacting neutrinos can be robustly approximated as a constant, $\ell_\infty=\mathcal{A}_\infty \times \ell^{\rm SM\nu}_\infty$, whose magnitude reflects the ``average'' neutrino propagation speed during the relevant cosmic epochs. For a more general scattering rate $\Gamma_\nu\propto T_\nu^n$, we demonstrate in Appendix~\ref{sec:A2} that a linear rescaling of $\ell^{\rm SM\nu}_\infty$ is a robust approximation as long as $n<7$, well above the physical models relevant for this work.

A larger temperature power law index $n$ means that the scattering rate changes more rapidly with redshift. As a consequence, for a given $z_{\nu,\rm dec}$, the steeper the temperature power law, the faster neutrinos approach their asymptotic free-streaming and fluid-like behavior at redshifts above and below $z_{\nu,\rm dec}$, respectively. As shown in Fig.~\ref{fig:3}, the $\Gamma_\nu \propto T_\nu^5$ case exhibits a faster, more sharply defined transition from the free-streaming to fluid-like phase shift regimes compared to $\Gamma_\nu \propto T_\nu^3$. More specifically, the neutrino-induced phase shift for $\Gamma_\nu \propto T_\nu^5$ reduces to the fluid-like (free-streaming) limit for $z_{\nu,\rm dec}\lesssim 10^{3}$  ($z_{\nu,\rm dec}\gtrsim10^{5}$); in contrast, the phase shift for $\Gamma_\nu \propto T_\nu^3$ reaches the fluid-like (free-streaming) limit for $z_{\nu,\rm dec}\lesssim 10^{2}$  ($z_{\nu,\rm dec}\gtrsim10^{6}$).

While we expect the fluid-like limit to be identical for both temperature dependencies, Fig.~\ref{fig:3} shows minor differences between the $\Gamma_\nu \propto T_\nu^3$ and $\Gamma_\nu \propto T_\nu^5$ scenarios at low $z_{\nu,\rm dec}$. Differences are also visible in Fig.~\ref{fig:2}: the template fits for the two scenarios have slightly different values as they approach the fluid-like limit. This apparent discrepancy arises from including the peaks and troughs of the TT power spectrum in our template fits. Compared to the EE and TE power spectra, the TT spectrum has a more complex transfer function, which makes it inherently difficult to cleanly isolate the phase shift~\cite{Montefalcone:2025unv}, in turn biasing the fitted amplitude. The fluid-like limits are indeed identical when fitting only the TE and EE peaks and troughs.
We discuss this issue further in Appendix~\ref{sec:A2} (see Fig.~\ref{fig:A1}). For self-consistency with the CMB data we utilize in our observational analyses, we use the full lensed power spectra fits throughout this work, noting that these minor differences have a negligible impact on our neutrino decoupling constraints. 

Finally, we note that fluid-like neutrinos induce a nonzero phase shift in the CMB power spectra with amplitude $\mathcal{A}_\infty \approx 0.3$ relative to that of SM free-streaming neutrinos, a result that appears to have been overlooked in previous studies. 
In Appendix~\ref{sec:A1}, we compare the spectrum-based templates for both free-streaming and fluid-like neutrinos (see Fig.~\ref{fig:A0}). The phase shift induced by three fluid-like neutrinos is roughly equivalent to that from approximately 0.5 free-streaming neutrinos. Fluid-like neutrinos still propagate faster than the photon-baryon fluid due to baryon inertia, allowing them to continue inducing a phase shift even when tightly coupled.

\section{Constraints from current CMB observations}\label{sec:3}

\noindent Having derived the spectrum-based templates that characterize the phase shift signatures of interacting neutrinos, we now incorporate these theoretical predictions into the analysis pipeline developed in Ref.~\cite{Montefalcone:2025unv}. We use this modified pipeline to constrain the neutrino-decoupling redshift for both the $\Gamma_\nu \propto T_\nu^3$ and $\Gamma_\nu \propto T_\nu^5$ interaction scenarios using the latest CMB data from \textit{Planck}, ACT, and SPT. 

\subsection{Analysis pipeline}

\noindent Our template function of Eq.~\eqref{eq:fell_int} introduces the asymptotic amplitude ratio $\mathcal{A}_\infty$ as a parameter that controls the magnitude of the phase shift relative to SM expectations, allowing us to probe deviations from the free-streaming neutrino behavior.
We implement the template by artificially introducing a multipole shift
\begin{align}
    \Delta\ell (\mathcal{A}_\infty,\Neff)\equiv \left(\mathcal{A}_\infty -1\right) \times A(\Neff)f^{\rm SM\nu}_\ell, \label{eq:deltaell_data}
\end{align}
where $f_\ell^{\rm SM\nu}$ represents the established spectrum template for SM free-streaming neutrinos~\cite{Montefalcone:2025unv}. This parametrization ensures that the standard physical scenario of $\Neff$ free-streaming neutrinos, corresponding to $\mathcal{A}_\infty = 1$, introduces no artificial shift in the power spectra.

By directly fitting for $\mathcal{A}_\infty$, we can robustly test for deviations from the SM through a signature-driven approach. The resulting model-agnostic measurements of $\mathcal{A}_\infty$ can then be mapped to constraints on the neutrino decoupling redshift using the numerically derived relationships established in the previous section (see Fig.~\ref{fig:3}), thereby providing limits on realistic neutrino interaction scenarios.

To implement Eq.~\eqref{eq:deltaell_data} in a CMB analysis, we closely follow the methodology of Refs.~\cite{Follin:2015hya, Montefalcone:2025unv}. The procedure involves first extracting the undamped power spectra by removing exponential diffusion damping [see Eq.~\eqref{eq:K_ell}], then decomposing the TT and TE power spectra into their integrated Sachs-Wolfe (ISW) and non-ISW components. The multipole shift $\Delta\ell$ is applied exclusively to the ISW-free component. The shifted power spectra are reconstructed by combining the modified non-ISW terms with the original ISW and cross-correlation components, and then we reapply the appropriate damping factors. The final result of this procedure is the shifted CMB power spectrum
\begin{equation}
    C_\ell\rightarrow \tilde{C}_\ell= C_\ell^{\rm ISW}+C_{\ell+\Delta\ell}^{\cancel{\rm ISW}}+C_\ell^{\rm cross}, \label{eq:deltaCell_data}
\end{equation}
where $\Delta\ell$ is given by Eq.~\eqref{eq:deltaell_data}. Here, we suppress the superscripts for the type of spectrum, and the decomposition of the power spectra into their ISW~and ISW-less components is only relevant for the TT~and TE~power spectra.

Our approach is effectively equivalent to the methodology developed in Ref.~\cite{Montefalcone:2025unv}, which introduced an effective number of multipole-shifting relativistic species, $\Neff^{\delta\ell}$, that exclusively controls the magnitude of the phase shift. As we demonstrate in Appendix~\ref{sec:A3}, constraints on $\Neff^{\delta\ell}$ can be directly mapped to constraints on our amplitude parameter $\mathcal{A}_\infty$. We adopt the $\mathcal{A}_\infty$ parametrization throughout this work, because it more explicitly connects the observed phase shift to the underlying neutrino interaction physics and, crucially, enables a direct mapping to neutrino decoupling redshifts through the numerically derived relationships presented in the previous section.

\subsection{Data analysis and constraints}

\begin{figure*}
       \centering
       \includegraphics[width=\linewidth]{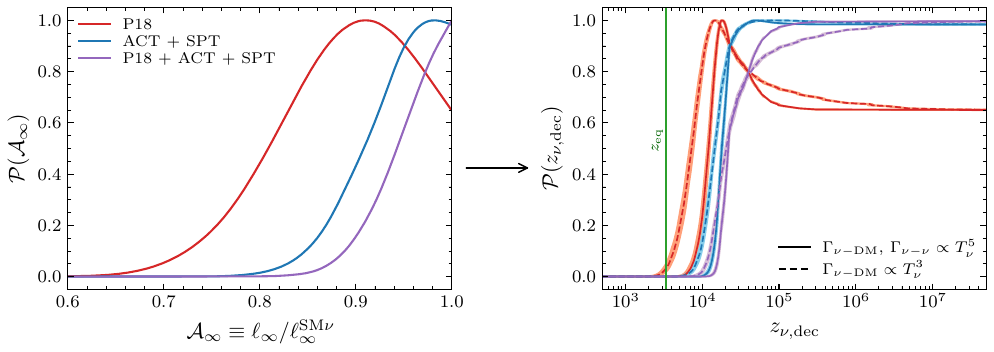}
    \caption{{\it Left:} One dimensional posterior distributions of the phase-shift-amplitude ratio $A_\infty$ for \textit{Planck} 2018 only (P18) in \textcolor{tabred}{red}, ACT and SPT data (ACT + SPT) in \textcolor{tabblue}{blue}, and the combined P18 + ACT + SPT analysis in \textcolor{tabpurple}{purple}. {\it Right:} Corresponding posterior distributions for the neutrino decoupling redshift $z_{\nu,\rm dec}$ for interacting neutrinos with scattering rates $\Gamma_\nu\propto T_\nu^3$ (dashed) and $\Gamma_\nu\propto T_\nu^5$ (solid). These posteriors are obtained by mapping the $\mathcal{A}_\infty$ constraints through the numerically derived $\mathcal{A}_\infty - z_{\nu,\rm dec}$ relationships shown in Fig.~\ref{fig:3}, with the shaded bands representing the propagated uncertainty in this mapping. The vertical \textcolor{tabgreen}{green} line marks the redshift of matter–radiation equality, $z_{\rm eq}$, in our fiducial $\Lambda$CDM model. All analyses strongly constrain neutrino interactions, requiring decoupling to occur deep within the radiation-dominated epoch.}
    \label{fig:4}
\end{figure*}

\noindent With our theoretical framework and analysis pipeline established, we now apply our methodology to constrain neutrino interactions using observed CMB data. We perform Markov chain Monte Carlo (MCMC) analyses using the publicly available samplers \texttt{MontePython}~\cite{Audren:2012wb, Brinckmann:2018cvx} and \texttt{Cobaya}~\cite{2019ascl.soft10019T,Torrado:2020dgo}, with a modified version of the Boltzmann code \texttt{CLASS} as in Ref.~\cite{Montefalcone:2025unv} that implements the prescription in Eqs.~\eqref{eq:deltaell_data} and~\eqref{eq:deltaCell_data}.

In all of our analyses, we use broad flat priors to sample over the six baseline parameters of the standard $\Lambda$CDM model: the angular size of the sound horizon at decoupling $\theta_s$, the physical baryon density $\omega_b$, the physical matter density $\omega_m$, the scalar amplitude $\ln(10^{10}A_s)$, the scalar spectral index $n_s$, and the photon optical depth $\tau_{\rm reio}$.
We infer the value of $\mathcal{A}_\infty$ while fixing $\Neff$ to its SM expectation $\Neff^{\rm SM} = 3.044$. For the phase shift parameter, we assume a flat prior in the range $\mathcal{A}_\infty \in[0, 1]$, which spans conservatively below the fluid-like limit up to the free-streaming SM scenario.\footnote{We note that our results are broadly insensitive to the specific choice of prior, as demonstrated by the agreement between our $\mathcal{A}_\infty$ results and those from the $\Neff^{\delta\ell}$ analysis presented in Ref.~\cite{Montefalcone:2025unv}. We provide a detailed comparison of the two approaches and their prior mappings in Appendix~\ref{sec:A3}.}

Chain convergence is reached when the Gelman-Rubin criterion~\cite{Gelman:1992zz} $R < 0.01$ is satisfied. We utilize \texttt{GetDist}~\cite{2019arXiv191013970L} to analyze our MCMC chains.
\smallskip

We consider the following datasets in our analyses:
\begin{itemize}
    \item \textit{Planck} 2018 (P18): We employ the \textit{Planck} 2018 PR3 likelihood code~\cite{Planck:2019nip}, specifically the \texttt{plik-lite} likelihood for high multipoles ($\ell>30$), and the \texttt{commander} and \texttt{SimAll} likelihoods for low-multipole ($2\leq \ell \leq 29$) TT and EE power spectra, respectively.
    \item ACT: We utilize the ACT DR6 \texttt{ACT-lite} likelihood~\cite{ACT:2025fju}, which provides TT, TE, and EE measurements for $\ell>600$. Due to the lack of large-scale data, for standalone ACT analyses, we follow Ref.~\cite{ACT:2025fju} and apply a Gaussian prior on the optical depth $\tau_{\rm reio}= 0.0566\pm 0.0058$. When combining ACT with \textit{Planck} data, we implement the ``$\it{Planck_{\rm cut}}$" dataset to avoid correlations, due to substantial sky overlap between the two surveys. Specifically, we only use \textit{Planck} high-$\ell$ data, $30\leq \ell < 1000$ in TT and $30\leq \ell < 600$ in TE/EE, and substitute in the \texttt{Sroll2} likelihood for low-$\ell$ polarization measurements~\cite{ACT:2025fju}.
    \item SPT: We use the SPT-3G D1 \texttt{SPT-lite} likelihood~\cite{SPT-3G:2025bzu}, containing TT, TE, and EE measurements for $\ell>400$. As with ACT, our standalone SPT analyses employ the same Gaussian optical depth prior. Due to minimal sky overlap with other experiments, SPT data can be directly combined with other datasets without multipole cuts~\cite{SPT-3G:2025zuh}.
\end{itemize}

We perform analyses focusing on three key data combinations: (1) P18 only, serving as our benchmark analysis; (2) ACT + SPT, testing the constraining power of ground-based experiments due to their enhanced sensitivity to higher multipole modes compared to \textit{Planck}, particularly for EE; and (3) P18 + ACT + SPT, combining all available CMB data to provide the most stringent bounds on neutrino interactions through precise measurements of the corresponding phase shift. In our combined dataset analyses, we do not consider the latest \textit{Planck} 2021 (P21) PR4 likelihood code based on the \texttt{NPIPE} data release~\cite{Planck:2020olo}.

As noted in Refs.~\cite{ACT:2025fju,ACT:2025tim}, the multipole truncation at $\ell< 1000$ when combining \textit{Planck} with ACT data reduces the sensitivity to the choice of \textit{Planck} likelihood. This consideration is even less relevant for our analysis, which primarily depends on the positions of acoustic peaks and troughs for $\ell \gtrsim 1000$, where the phase shift is substantial.

For consistency, we present only P18 results in the main text. We also perform a standalone P21 analysis to verify consistency with P18, partly motivated by minor differences in phase shift measurements between these datasets identified in Ref.~\cite{Montefalcone:2025unv}. This P21 analysis and all supplementary data combinations are presented in Appendix~\ref{sec:A4}.

Finally, it is important to note that our analyses employ the compressed \texttt{lite} likelihoods for the three CMB experiments under consideration, which have parameters describing foregrounds and residual systematics already marginalized over. We have verified that this choice does not impact our results by explicitly comparing constraints on $\mathcal{A}_\infty$ obtained using both the \texttt{lite} and \texttt{full} likelihoods for each dataset, finding excellent agreement in all cases.

\subsubsection{Constraints on universal neutrino interactions}

\begin{table}
	\centering
	\begin{tabular}{l c @{\hskip 2em} c @{\hskip 1em} c}
			\toprule
			& & \multicolumn{2}{c}{$z_{\nu,\rm dec}$}\\
            \cmidrule(lr){3-4}
	Dataset & {$\mathcal{A}_\infty$} & {$\Gamma_\nu\propto T_\nu^3$} & {$\Gamma_\nu\propto T_\nu^5$} \\
            \midrule[0.065em]
	P18 & $>0.76$	& $> 7.9\times 10^3$	& $>1.27\times 10^4$	\\[4pt]
        ACT + SPT & $>0.87$	& $> 1.06\times 10^4$	& $>1.51\times 10^4$	\\[4pt]
        P18 + ACT + SPT & $>0.90$	& $> 1.33\times 10^4$	& $>1.71\times 10^4$ \\[2pt]
	\bottomrule
	\end{tabular}
	\caption{95\% C.L. lower limits on the phase-shift-amplitude ratio $\mathcal{A}_\infty$ and the corresponding neutrino decoupling redshifts $z_{\nu,\rm dec}$ for two interaction scenarios with scattering rates $\Gamma_\nu \propto T_\nu^3$ and $\Gamma_\nu \propto T_\nu^5$, respectively. Results are provided for three data combinations: \textit{Planck} 2018 only (P18), ACT and SPT combined (ACT + SPT), and all datasets combined (P18 + ACT + SPT). The decoupling redshifts are derived by mapping the $\mathcal{A}_\infty$ constraints through the numerically derived $\mathcal{A}_\infty - z_{\nu,\rm dec}$ relationships presented in Section~\ref{sec:2.2}.}
	\label{tab:1}
\end{table}

\noindent We now present the main results of our analysis, constraining universal neutrino interactions through measurements of the phase shift in CMB power spectra. Table~\ref{tab:1} summarizes the 95\% confidence level (C.L.) limits on $\mathcal{A}_\infty$ for our three primary data combinations, along with the corresponding neutrino decoupling redshifts for both interaction scenarios under consideration, i.e.\ for $\Gamma_\nu\propto T_\nu^3$ and $\Gamma_\nu\propto T_\nu^5$. Figure~\ref{fig:4} complements these results by displaying the one-dimensional posteriors for $\mathcal{A}_\infty$ (left panel) and the mapped posteriors for $z_{\nu,\rm dec}$ in both neutrino interaction models (right panel). The bands around the redshift posteriors reflect the propagated uncertainty from the numerically computed $\mathcal{A}_\infty - z_{\nu,\rm dec}$ relationship for each neutrino-interaction scenario, shown in Fig.~\ref{fig:3}.

Overall, our analyses unequivocally detect the neutrino-induced phase shift in the observed CMB data, consistent with previous studies~\cite{Follin:2015hya,Baumann:2015rya, Brust:2017nmv, Choi:2018gho,Kreisch:2019yzn, Forastieri:2019cuf, Blinov:2020hmc, Das:2020xke, Brinckmann:2020bcn, Saravanan:2025cyi, Montefalcone:2025unv}. Across all dataset combinations, we find that CMB measurements are consistent with the SM expectation of three free-streaming neutrinos, i.e.\ $\mathcal{A}_\infty = 1$, at the $1\sigma$ level. The only dataset whose posterior peaks marginally away from unity is the \textit{Planck}-only analysis, which nonetheless remains consistent with the SM expectation at the $\sim 1.1\sigma$ level, yielding $\mathcal{A}_\infty = 0.89^{+0.10}_{-0.05}$ at 68\% C.L.\ and $\mathcal{A}_\infty > 0.76$ at 95\% C.L.

As anticipated, the inclusion of ground-based experiments proves beneficial for more precise phase-shift measurements. Interestingly, we find that combined ACT and SPT data surpass \textit{Planck}'s constraining power on the phase shift. This enhanced sensitivity stems from improved measurements at higher multipoles, particularly in polarization power spectrum, where the ACT + SPT combination exceeds \textit{Planck}'s precision for $\ell > 600$ in EE and $\ell > 1000$ in TE~\cite{ACT:2025fju,SPT-3G:2025bzu}.  As further highlighted in Appendix~\ref{sec:A4}, ACT provides the dominant constraining power among ground-based experiments, demonstrating that most of the sensitivity to the phase shift arises from multipoles $\ell\lesssim 2000$, where ACT measurements are the most precise to date for both the EE and TE power spectra.  The dominance of intermediate multipoles in phase shift measurements is expected, as the exponential damping of acoustic oscillations gradually diminishes the extractable information on the phase shift at higher multipoles. In addition, we emphasize that the \textit{Planck}-inspired Guassian prior on the optical depth, $\tau_{\rm reio}$, applied in the ACT and SPT analyses does not affect the phase-shift constraint and therefore does not implicitly reintroduce \textit{Planck}'s sensitivity. In fact, at the multipoles relevant for the phase shift ($\ell \gg30$), $\tau_{\rm reio}$ enters only as an overall amplitude rescaling of the CMB power spectra, which is uncorrelated with the phase information. The primary contribution of \textit{Planck} to the joint analysis instead arises from its precise temperature measurements, which in particular lead to a more accurate determination of the angular size of the sound horizon $\theta_s$. Since $\theta_s$ is effectively the only $\Lambda$CDM parameter degenerate with $\mathcal{A}_{\infty}$~\cite{Montefalcone:2025unv}, this improved measurement directly enhances the phase-shift constraint (see also Appendix~\ref{sec:A4} for more details).

Combining all available CMB data yields the tightest constraints on neutrino interactions derived exclusively from phase shift measurements. We obtain $\mathcal{A}_\infty > 0.90$ at 95\% C.L., which translates to lower bounds on the neutrino decoupling redshift of $z_{\rm dec} > 1.33 \times 10^4$ and $z_{\rm dec} > 1.71 \times 10^4$ for $\Gamma_\nu \propto T_\nu^3$ and $\Gamma_\nu \propto T_\nu^5$ interactions, respectively. These redshift bounds correspond to times significantly before matter-radiation equality, indicating that current CMB data require neutrinos to be free-streaming deep within the radiation-dominated epoch. Notably, for $\Gamma_\nu \propto T_\nu^5$, our phase shift-only constraint is comparable with the limit obtained from a full CMB analysis of SI neutrinos that include both phase shift and amplitude effects, yielding $z_{\rm dec} \gtrsim 10^5$ at 95\% C.L.~\cite{RoyChoudhury:2020dmd,Poudou:2025qcx}. These bounds further demonstrate that the phase shift not only serves as a targeted probe of neutrino self-interactions, but it alone can approach a similar level of sensitivity to a complete analysis.

\begin{figure}[t!]
    \centering
    \includegraphics[width=\linewidth]{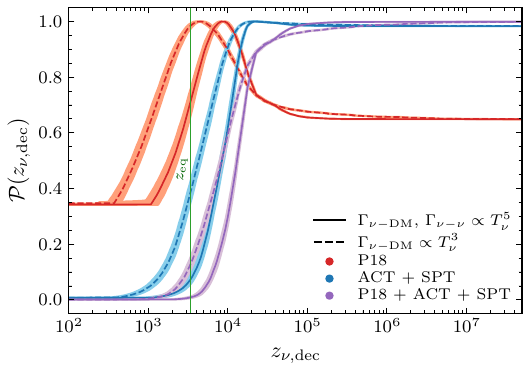}
    \caption{Same as the right panel of Fig.~\ref{fig:4}, but for the flavor-dependent scenario, in which only one neutrino species is interacting, i.e.\ $\mathcal{F}_{\nu, \rm int}=1/3$. These posteriors are obtained by first rescaling the constraints on the universal amplitude ratio $\mathcal{A}_\infty$ to the flavor-dependent ratio $\mathcal{A}'_\infty$ via Eq.~\eqref{eq:Aprime_infty}, then converting to $z_{\nu,\rm dec}$ through the same numerically derived $A_\infty -z_{\nu,\rm dec}$ relationships shown in Fig.~\ref{fig:3}. Restricting interactions to a single neutrino species significantly weakens the constraints: P18 data permit fluid-like behavior extending to low redshifts, while combined datasets constrain neutrino decoupling to occur in the radiation-dominated era.}
    \label{fig:5}
\end{figure}

\subsubsection{Constraints on flavor-dependent neutrino interactions}

\noindent It is possible that neutrino interactions exhibit flavor dependence, for which only a subset of neutrino species have non-standard interactions~\cite{Das:2020xke,Brinckmann:2020bcn,Das:2023npl,RoyChoudhury:2022rva}. Such flavor-dependent scenarios have been explored in the literature, particularly in the context of neutrino self-interactions, and CMB data can accommodate stronger couplings when fewer flavors are involved~\cite{Das:2020xke}.

Since the phase shift amplitude scales linearly with the fractional energy density in neutrinos, we can straightforwardly extend our framework to constrain these more general interaction models. For a scenario in which only a fraction $\mathcal{F}_{\nu,\rm int}$ of neutrinos participate in interactions, the corresponding asymptotic amplitude ratio for this flavor-dependent case, $\mathcal{A}'_\infty$, relates to our measured universal amplitude $\mathcal{A}_\infty$ according to
\begin{equation}
(1 - \mathcal{A}'_\infty) = (1 - \mathcal{A}_\infty)/\mathcal{F}_{\nu,\rm int}. \label{eq:Aprime_infty}
\end{equation}
As expected, $\mathcal{A}'_\infty = \mathcal{A}_\infty$ when all neutrinos interact ($\mathcal{F}_{\nu,\rm int} = 1$). Generically for $\mathcal{F}_{\nu,\rm int}<1$, $\mathcal{A}'_\infty< \mathcal{A}_\infty$, corresponding to a more strongly coupled scenario; if only a subset of neutrinos can interact, they must interact more strongly to produce the same observable phase shift as three weakly interacting species.
 
Given this relationship, we can map the constraints on $\mathcal{A}_{\infty}$ presented in the previous subsection to the corresponding limits on $\mathcal{A}^\prime_{\infty}$ for any specified value of $\mathcal{F}_{\nu,\rm int}$. The resulting $\mathcal{A}^\prime_{\infty}$ bounds can then be translated to a decoupling redshift using the same numerically computed $\mathcal{A}_\infty - z_{\nu,\rm dec}$ relationships established in Section~\ref{sec:2.2}. The $\mathcal{A}_\infty - z_{\nu,\rm dec}$ mapping remains valid for $\mathcal{F}_{\nu,\rm int} < 1$, as it simply characterizes the overall scaling of the phase-shift template $f_\ell$ for different neutrino interaction scenarios.

We illustrate the impact of flavor-dependent interactions in Fig.~\ref{fig:5} for the limiting case $\mathcal{F}_{\nu,\rm int} = 1/3$, corresponding to only one neutrino species participating in interactions.\footnote{Note that for a given observed $\mathcal{A}_\infty<1$, choosing too small a value of $\mathcal{F}_{\nu,\rm int}$ can push $\mathcal{A}^\prime_\infty$ into unphysical values below the fluid-like limit ($\sim0.3$), indicating a phase shift smaller than is physically realizable for the assumed interacting fraction. In such cases, the resulting $z_{\nu,\rm dec}$ constraints may be skewed by effectively discarding part of the original posterior volume. This complication does not arise in our analysis with $\mathcal{F}_{\nu,\rm int}=1/3$,  since the tight constraints on $\mathcal{A}_\infty$ in the universal interaction scenario keep the mapped values of $\mathcal{A}^\prime_\infty$ within the physical regime. } As expected, restricting interactions to a single species significantly weakens the constraints on $z_{\nu,\rm dec}$. The P18-only analysis shows that the data permit fluid-like behavior extending to low redshifts, with a broad posterior peak around $z_{\nu,\rm dec} \sim 5\times 10^3$ and $ \sim 9\times 10^3$ for $\Gamma_\nu\propto T_\nu^3$ and $\Gamma_\nu\propto T_\nu^5$, respectively. Our P18-only analysis for the $\Gamma_\nu\propto T_\nu^5$ case provides an independent confirmation of the findings in Ref.~\cite{Das:2020xke}, which showed that \textit{Planck} data is compatible with one strongly-interacting neutrino flavor that has a late decoupling $z_{\nu , \rm dec} \approx 8.8\times 10^3$ just prior to matter-radiation equality. The broad agreement between that dedicated analysis and our signature-driven approach highlights the robustness of using the phase shift as a probe of neutrino interactions.

When we include ground-based experiments, either alone or in combination with \textit{Planck}, we obtain constraints on the decoupling redshift, though less stringent than in the universal interaction case. With the sole exception of the $\Gamma_\nu \propto T_\nu^3$ scenario analyzed with ACT + SPT alone, all constraints require neutrino decoupling to occur before matter-radiation equality. The tightest bounds come from the combined P18 + ACT + SPT analysis, which constrains $z_{\nu,\rm dec} > 7.3 \times 10^3$ and $z_{\nu,\rm dec} > 3.6 \times 10^3$ for $\Gamma_\nu \propto T_\nu^5$ and $\Gamma_\nu \propto T_\nu^3$ at 95\% C.L., respectively. These results demonstrate that even in flavor-dependent scenarios, current CMB data provide robust constraints on neutrino interactions from direct phase-shift measurements.

\section{Discussion \& Summary}\label{sec:4}

\noindent In this work, we demonstrate that the phase shift in the acoustic oscillations of the CMB provides a powerful direct probe of neutrino interactions in the early universe. By following the full evolution of neutrino perturbations in realistic interaction models with temperature-dependent scattering rates, we show how neutrino interactions modify this characteristic phase shift in ways that can be robustly constrained with current data.

Our theoretical analysis reveals that the neutrino-induced phase shift is well-described by the same functional form as in the free-streaming case, but scaled by the asymptotic amplitude $\mathcal{A}_\infty$, which varies smoothly as a function of decoupling redshift. The amplitude ranges from $\mathcal{A}_\infty = 1$ (corresponding to the free-streaming limit) to $\mathcal{A}_\infty \approx 0.3$ (corresponding to the fluid-like regime).

We explicitly quantify for the first time that even fluid-like neutrinos, which never fully decouple to allow free-streaming, induce a nonzero phase shift compared to the free-streaming case in the SM. The sound speed of fluid-like neutrinos remains higher than that of the photon–baryon fluid due to baryon inertia.

We apply our framework to constrain two classes of neutrino-temperature-dependent interaction rates: $\Gamma_\nu \propto T_\nu^3$ and $\Gamma_\nu \propto T_\nu^5$, which correspond to models of neutrino-DM scattering~\cite{Boehm:2003hm,Stadler:2019dii,Serra:2009uu,Olivares-DelCampo:2017feq,Ghosh:2017jdy,Ghosh:2024wva,Wilkinson:2014ksa,Escudero:2015yka,DiValentino:2017oaw,Ghosh:2019tab,Mosbech:2020ahp} and neutrino self-interactions~\cite{Cyr-Racine:2013jua,Oldengott:2014qra,Oldengott:2017fhy,Kreisch:2019yzn,Das:2020xke,Das:2023npl,RoyChoudhury:2020dmd,Lancaster:2017ksf,Loverde:2022wih, Camarena:2023cku,Poudou:2025qcx,He:2023oke,He:2025jwp}. Using the latest CMB data from \textit{Planck}, ACT, and SPT, we constrain the neutrino decoupling redshift using phase-shift measurement, firmly establishing that neutrinos must have been freely streaming since deep within the radiation-dominated era. 
We also explore a flavor-dependent scenario in which only one neutrino species participates in the interaction. We find that even in this limiting scenario, combined CMB data still require neutrino decoupling to occur in the radiation-dominated era; however, our \textit{Planck}-only analysis for the $\Gamma_\nu \propto T_\nu^5$ scenario permits fluid-like behavior extending to low redshifts.

Current ground-based CMB experiments surpass \textit{Planck}'s sensitivity to the neutrino phase shift. The combination of ACT and SPT data alone provides stronger constraints on the neutrino decoupling redshift than \textit{Planck}, driven by more precise measurements of the EE polarization power spectrum at high multipoles, where the phase-shift signal-to-noise ratio peaks.
Upcoming data from the Simons Observatory~\cite{SimonsObservatory:2025wwn} will enable unprecedented precision in measuring the neutrino-induced phase shift~\cite{Montefalcone:2025unv}. Moreover, CMB constraints could be further strengthened by combined analyses with large-scale structure datasets~\cite{Baumann:2017gkg,Baumann:2019keh,Whitford:2024ecj}, since the same phase shift is imprinted in the baryon acoustic oscillations~\cite{Baumann:2017lmt,Baumann:2017gkg,Green:2020fjb, Montefalcone:2025mbg}. While existing forecasts indicate that CMB measurements are expected to dominate the overall sensitivity~\cite{Baumann:2017gkg,Baumann:2019keh,Whitford:2024ecj}, incorporating BAO information provides an independent probe of the same neutrino-induced phase shift, enabling a complementary consistency check and further reducing residual parameter degeneracies. In practice, this would require modeling the impact of neutrino interactions on the phase shift imprinted in the BAO feature, analogously to the CMB-based study presented here.

As a whole, our framework builds on previous literature~\cite{Follin:2015hya, Baumann:2015rya, Pan:2016zla, Baumann:2017gkg, Baumann:2019keh,Ghosh:2019tab,Brinckmann:2020bcn,Ge:2022qws,Taule:2022jrz,Ghosh:2024wva,Whitford:2024ecj,Montefalcone:2025unv, Saravanan:2025cyi, Montefalcone:2025mbg} investigating neutrino interactions through the phase shift of acoustic oscillations in the CMB. We introduce a model-agnostic way to search for deviations from the SM expectation of three free-streaming neutrinos, while maintaining a direct connection to the underlying physics through the numerically derived relationship between $\mathcal{A}_\infty$ and the decoupling redshift of neutrinos, within a given interaction model. While our focus has been on neutrinos, the same methodology is also applicable to dark radiation with similar interactions, and it could naturally be generalized to broader classes of dark radiation scenarios~(see e.g.~\cite{Cyr-Racine:2012tfp,Forastieri:2019cuf,Bansal:2021dfh,Ghosh:2021axu,Brinckmann:2022ajr,Sandner:2023ptm,Bansal:2022qbi,Das:2025asx}) that can imprint characteristic modifications to the phase-shift template. 

Several extensions to this work merit future investigation. For example, models with very high temperature scalings or resonant interactions~\cite{Creque-Sarbinowski:2020qhz,Venzor:2023aka,Noriega:2025ulc} would produce scale-dependent modifications to the phase shift, requiring the development of new phase-shift templates. Additionally, DM loading in neutrino-DM scattering models~\cite{Ghosh:2024wva} induces a phase shift; interactions with DM can slow the neutrino fluid below the sound speed of the primordial plasma, potentially reversing the direction of the phase shift. Finally, we note that incorporating the perturbation-based template introduced in Ref.~\cite{Montefalcone:2025unv} could enable more robust analyses of beyond-SM scenarios. We leave explorations of these possibilities to future work.

\acknowledgments
We thank Benjamin Wallisch for the valuable discussions and insightful comments on a preliminary version of this manuscript. KB and SG acknowledge support from the National Science Foundation under Grant No.~PHY-2413016. DH and YT were partially supported by the NSF Grant PHY-2412701. YT would like to thank the Tom and Carolyn Marquez Chair Fund for its generous support and would also like to thank the host of Fermilab, supported by the URA-22-F-13 fund. KB and YT would like to thank the Aspen Center for Physics (supported by NSF grant PHY-2210452). We acknowledge the use of \texttt{CLASS}~\cite{Blas:2011rf}, \texttt{Cobaya}~\cite{2019ascl.soft10019T,Torrado:2020dgo}, \texttt{GetDist}~\cite{2019arXiv191013970L}, \texttt{IPython}~\cite{Perez:2007ipy} and  \texttt{MontePython}~\cite{Audren:2012wb, Brinckmann:2018cvx}, and the Python packages \texttt{Matplotlib}~\cite{Hunter:2007mat}, \texttt{NumPy}~\cite{Harris:2020xlr}, and~\texttt{SciPy}~\cite{Virtanen:2019joe}. We acknowledge the Texas Advanced Computing Center (TACC) at The University of Texas at Austin for providing high-performance computing resources that have contributed to the research results reported within this
paper.

\appendix
\section{\texttt{CLASS}-based implementation of neutrino interactions}\label{sec:A1}

\noindent In this appendix, we describe the implementation of neutrino interaction models in the Boltzmann solver \texttt{CLASS} that we use to compute the phase-shift templates presented in the main text. We employ two distinct numerical frameworks, depending on the interaction type: for neutrino-DM scattering, we utilize the built-in interacting dark matter-dark radiation (\texttt{idm\_idr}) module~\cite{Archidiacono:2019wdp}, while for neutrino self-interactions, we use \texttt{nuCLASS}~\cite{Libanore:2025ack}, a modified version of \texttt{CLASS} specifically designed to handle neutrino-neutrino scattering.

Throughout our calculations, we ensure numerical accuracy by using stringent precision settings, particularly for the neutrino perturbation hierarchy, which is crucial for accurate phase shift extraction. We verify that for a given temperature scaling of the interaction rate and the same decoupling redshift, both implementations yield identical phase shift signatures, confirming the universality of our results.

\subsection{Neutrino-dark matter scattering}
\noindent For neutrino-DM interactions, we utilize the built-in \texttt{idm\_idr} module  in \texttt{CLASS}, which is based on the ETHOS framework~\cite{Cyr-Racine:2015ihg}. We treat all neutrinos as interacting dark radiation by setting $\texttt{N\_ur}=0$  (no free-streaming relativistic species) and $\texttt{N\_idr}=N_{\rm eff}$ for the desired effective number of relativistic neutrino species. 

The comoving scattering rate $\gamma$ between neutrinos and DM is parametrized as
\begin{equation}
    \gamma_{\nu-\rm iDM}=f_{\rm iDM}\, \omega_{c}\,a_{\rm dark}\,\left(\frac{1+z}{1+z_d}\right)^{n_{\rm dark}}, \label{eq:A1}
\end{equation}
where $f_{\rm iDM}$ (\texttt{f\_idm})  is the fraction of interacting DM, $\omega_c$ is the total physical cold DM density, $z_d=10^7$ is a normalization redshift, and $n_{\rm dark}$ (\texttt{nindex\_dark}) and $a_{\rm dark}$ (\texttt{a\_dark}) control the temperature dependence and overall strength of the interaction, respectively. A given value of $n_{\rm dark}$ corresponds to a physical interaction rate $\Gamma_{\nu-\rm DM}\propto T_\nu^{n_{\rm dark} +1}$.  Thus, to model the $\Gamma_{\nu-\rm DM} \propto T_\nu^3$ and $\Gamma_{\nu-\rm DM} \propto T_\nu^5$ scenarios discussed in the main text, we set $n_{\rm dark} = 2$ and $n_{\rm dark} = 4$, respectively. To isolate the phase shift signature, we set $f_{\rm iDM} = 10^{-3}$ throughout our calculations. This small fraction ensures that the modified propagation speed of neutrino perturbations---and hence the phase shift---is the only observable effect on the CMB, while dark acoustic oscillations and DM loading~\cite{Ghosh:2024wva} remain negligible. For a given target decoupling redshift $z_{\nu,\rm dec}$, we tune the interaction strength $a_{\rm dark}$ such that the condition    $\gamma_{\nu-\rm iDM}(a_{\rm dark},z_{\nu,\rm dec})\cdot (1+z_{\nu,\rm dec})/H(z_{\rm \nu,dec})=1$ is satisfied, corresponding to the time when the scattering rate falls below the Hubble expansion rate.

To properly account for neutrino interaction effects in the numerical implementation, the \texttt{idm\_idr} module requires specifying an additional parameter: \texttt{idr\_nature}. Setting this parameter to \texttt{free-streaming} evolves the full Boltzmann hierarchy for the neutrino perturbations up to a maximum multipole $\texttt{l\_max\_idr}$ (which we set to 50 throughout), capturing the complete evolution from tightly coupled to free-streaming behavior. The alternative \texttt{fluid} setting assumes perpetually tightly coupled neutrinos, evolving only the first two moments of the Boltzmann hierarchy~\cite{Archidiacono:2019wdp}. We have verified that our implementation correctly reproduces both limiting cases: SM free-streaming neutrinos in the limit of negligible interaction strength and fully fluid-like behavior for very strong interactions.  Finally, since we utilize this implementation only to describe neutrino-DM interactions without neutrino self-interactions, we also set the parameter $\texttt{b\_idr}=0$ throughout.

\subsection{Neutrino Self-Interactions}

\noindent For neutrino self-interactions, we utilize \texttt{nuCLASS}~\cite{Libanore:2025ack}, which implements a phenomenological parametrization of neutrino-neutrino scattering, following the framework of Ref.~\cite{Kreisch:2019yzn}. In this approach, the low-energy effects on cosmology are captured through an effective four-neutrino interaction, characterized by a dimensionful Fermi-like coupling constant $G_{\rm eff}$, yielding a physical interaction rate
\begin{equation}
    \Gamma_{\nu-\nu} = G_{\rm eff}^2 T_\nu^5 .\label{eq:A2}
\end{equation}
Similar to the neutrino-DM scattering scenario, we treat all neutrinos as interacting by setting $\texttt{N\_ur}=0$, $\texttt{N\_ncdm}=1$ and adjusting $\texttt{deg\_ncdm}$ to produce the desired value of $\Neff$. For a given target decoupling redshift $z_{\nu,\rm dec}$, we tune the coupling constant $G_{\rm eff}$ such that $\Gamma_{\nu-\nu}(G_{\rm eff},z_{\nu,\rm dec}) = H(z_{\nu,\rm dec})$.

The \texttt{nuCLASS} implementation evolves the full Boltzmann hierarchy for self-interacting neutrinos up to a maximum multipole $\texttt{l\_max\_idr}$, which we again set to 50 to accurately account for the transition from the tightly coupled regime at high redshifts to the free-streaming regime after decoupling. We have verified that this implementation produces phase-shift templates identical to those obtained with the \texttt{idm\_idr} module for the $\Gamma_{\nu-\rm DM} \propto T_\nu^5$ case at equivalent decoupling redshifts. This agreement confirms that the phase shift depends only on the temperature scaling and decoupling epoch, not on the specific interaction mechanism.

\section{Phase-shift template extraction and validation}\label{sec:A2}

\begin{table}
	\centering
	\sisetup{group-digits=false}
	\begin{tabular}{l S[table-format=1.5]}
			\toprule
		Parameter 				& {Fiducial Value}															\\
			\midrule[0.065em]
		$\omega_b$ 				& 0.02238 				 		\\
		$\omega_c$				& 0.12011					\\
		$100\,\theta_s$ 		& 1.04178 						\\
		$\ln(\num{e10}A_s)$		& 3.0448							\\
		$n_s$					& 0.96605 													\\
		$\tau$ 					& 0.0543 									\\
		$N_{\rm eff}$					& 3.044 							\\
		$Y_p$					& {`BBN'\hskip6.5pt}\\
    
			\bottomrule
	\end{tabular}
	\caption{Fiducial $\Lambda$CDM parameters adopted in this work, based on the {\it Planck 2018} best-fit values~\cite{Planck:2018vyg}. The parameters are the physical baryon density $\omega_b$; the cold dark matter density $\omega_c$; the angular size of the sound horizon $\theta_s$; the amplitude of primordial scalar perturbations, $\ln(10^{10}A_s)$, and spectral index, $n_s$, at the pivot scale $k_\star = 0.05\,{\rm Mpc}^{-1}$; the optical depth due to reionization $\tau$; the effective number of relativistic neutrino species $N_{\rm eff}$; and the primordial helium fraction $Y_p$. We treat neutrinos as massless throughout, which is an excellent approximation for phase-shift extraction purposes~\cite{Montefalcone:2025unv}. $Y_p$ is generally fixed by requiring consistency with Big Bang Nucleosynthesis, which yields $Y_p = 0.24534$ for this set of parameters.}
\label{tab:parameters}
\end{table}

\noindent In this appendix, we provide technical details underlying our phase shift analysis of interacting neutrinos. We present our generalized method for extracting phase-shift templates from CMB power spectra: we adapt existing  techniques used in the context of free-streaming neutrinos to robustly handle the full range of neutrino interaction scenarios. We then validate the key approximation used throughout the main text: the phase shift from interacting neutrinos can be parametrized by a simple amplitude rescaling of the SM template. Our validation demonstrates that this approximation holds for all physically motivated interaction models and quantifies minor systematic effects from including the temperature power spectrum, confirming the robustness of our analysis pipeline.

\subsection{Generalized phase shift extraction from CMB power spectra}
\noindent In this section, we discuss our generalized procedure used to extract the phase shift from CMB power spectra. We first briefly review the method of Ref.~\cite{Follin:2015hya,Montefalcone:2025unv}, developed for the case of free-streaming neutrinos. We then describe the modifications introduced here to improve accuracy and stability in the presence of neutrino interactions.

\subsubsection{Free-streaming neutrinos}

\begin{figure}
    \centering
    \includegraphics[width=\linewidth]{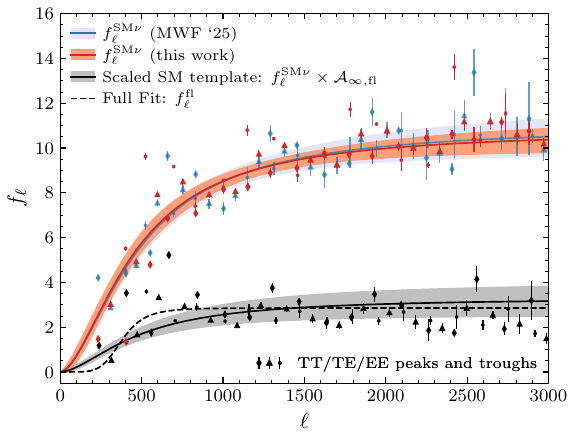}
    \caption{Comparison of the phase-shift templates, Eq.~\eqref{eq:fell}, for SM free-streaming neutrinos obtained in Ref.~\cite{Montefalcone:2025unv} (\textcolor{tabblue}{blue}) and this work (\textcolor{tabred}{red}).  Diamonds, triangles, and circles represent the peaks and troughs of the TT, TE, and EE spectra respectively. The best-fit templates with corresponding $2\sigma$ confidence intervals are displayed in their respective colors, showing remarkable consistency between the two methods with slightly reduced uncertainties at higher multipoles for our two-step approach. In black, we also show the template for fluid-like neutrinos. We compare the best fit obtained by fitting only the amplitude ratio $\mathcal{A}_\infty$ multiplied by the SM template, Eq.~\eqref{eq:fell_int}, (solid line with shaded confidence region) against the full free fit where all three template parameters vary, Eq.~\eqref{eq:fell} (dashed line). The good agreement between these two fluid-like fits, especially at high multipoles, validates our amplitude-rescaling parametrization that we use throughout the main text.}
    \label{fig:A0}
\end{figure}

\noindent In the free-streaming case, the phase-shift template, as defined in Eq.~\eqref{eq:delta_ell}, is obtained by numerically computing the multipole shifts $\delta\ell$ from the lensed CMB spectra generated with \texttt{CLASS}, ensuring self-consistency with the observed power spectra, which are also lensed.  Specifically, the procedure compares a fiducial $\Lambda$CDM model (see parameters in Table~\ref{tab:parameters}) with 100 cosmologies of varying $N_{\rm eff}$ in the range $[1,6]$. The physical baryon density~$\omega_b$, the scale factor at matter–radiation equality $a_{\mathrm{eq}} \equiv \omega_m/\omega_r$,
and the angular size of the sound horizon~$\theta_s$ are held fixed to remove the effects on the oscillation frequency and radiation driving through changes in~$a_{\rm{eq}}$. The remaining background effect on the angular size of the Silk damping scale~$\theta_d$, from changes in the expansion rate due to different radiation densities, is also eliminated by adjusting the primordial helium fraction~$Y_p$, accordingly.

To further isolate the acoustic oscillations, we remove the ISW contributions to avoid contamination from post-recombination effects and approximately undo 
Silk damping, which would otherwise obscure the precise locations of the acoustic peaks and troughs. This last step in particular is achieved by defining the undamped, lensed power spectrum as~\cite{Follin:2015hya, Wallisch:2018rzj, Montefalcone:2025unv}
\begin{equation}
    \mathcal{K}_\ell^{XY} \equiv \frac{\ell(\ell+1)}{2\pi}\, C_\ell^{XY} \exp\left\{a(\theta_d\ell)^\kappa \right\},
    \label{eq:K_ell}
\end{equation}
where the parameters $a \approx 0.68$ and $\kappa \approx 1.3$ are obtained from a fit to the fiducial model for fixed $\theta_d \approx 1.6\times 10^{-3}$, following Ref.~\cite{Wallisch:2018rzj}.

Applying this procedure yields the SM phase-shift template $f^{\rm SM\nu}_\ell$ we refer to in the main text, with best-fit parameters $\ell^{\rm SM\nu}_\infty = 11.0 \pm 0.6$, $\ell^{\rm SM\nu}_\star = 483 \pm 53$, and $\xi^{\rm SM\nu} = -1.69 \pm 0.13$. The resulting template, together with its $2\sigma$ confidence interval, is shown in blue in Fig.~\ref{fig:A0}.

\subsubsection{Interacting neutrinos}

\noindent While robust, the phase shift isolation from the procedure described above is inherently only approximate, which is why Refs.~\cite{Follin:2015hya,Montefalcone:2025unv} sample over $\mathcal{O}(100)$ cosmologies with varying $\Neff$ to average over residual cosmology-dependent effects and determine a stable template shape. In the context of interacting neutrinos that decouple at a redshift $z_{\nu, \mathrm{dec}}$, this direct method becomes significantly more challenging to implement numerically. In addition to varying $\Neff$ and the other cosmological parameters across the 100 sampled models, one must also adjust the interaction strength in each case to ensure the same $z_{\nu, \mathrm{dec}}$ for all values of $\omega_{\mathrm{cdm}}$ and $\Neff$, within the underlying cosmologies.

To overcome these difficulties, we adopt a two-step approach. First, we compute the phase-shift template for fully {\it fluid-like} neutrinos using the same procedure reviewed above. Then, for a given interacting-neutrino model with specified decoupling redshift $z_{\nu,\rm dec}$, we determine the {\it relative} phase shift between this model and the corresponding fluid-like case, for each of the underlying 100 cosmologies with varying $N_{\rm eff}$. The final phase shift is obtained by summing the fluid-like template and the measured relative shift. In this formulation, all systematic uncertainty in the final $\delta\ell$ effectively originates from the fluid-like template extraction. In fact, the relative-shift calculation is practically exact, as the only difference between the two spectra being compared is the effect of the interactions on neutrino perturbations. This procedure yields improved accuracy and numerical stability, particularly for scenarios in which neutrinos are in an intermediate regime between free-streaming and fully fluid-like behavior.

We validate this method by recomputing the SM free-streaming template, shown in red in Fig.~\ref{fig:A0} together with its $2\sigma$ confidence interval. The $\delta\ell$ values obtained through the two-step procedure are in excellent agreement with those from the direct extraction, with slightly smaller uncertainties at high multipoles, resulting in an effectively identical best-fit template.  In the same figure, we also display the numerical $\delta\ell$ shifts for fully fluid-like neutrinos, illustrating that they imprint a nonzero phase shift in the CMB power spectra, which approaches a constant value at large $\ell$, approximately one-third of the free-streaming amplitude. By rescaling $\Neff$, we find that the same phase shift can be produced by roughly $0.5$ effective free-streaming neutrinos, representing a non-negligible contribution to the CMB power spectra.

This reduced but nonzero phase shift in the fluid limit provides a natural baseline for understanding intermediate interaction scenarios. In Fig.~\ref{fig:A0}, we show in black the best-fit template and $2\sigma$ confidence region obtained by fitting only the asymptotic amplitude ratio $\mathcal{A}_\infty\equiv \ell_\infty/\ell_{\infty}^{\rm SM\nu}$, see Eq.~\eqref{eq:fell_int}, while keeping $\ell_\star$ and $\xi$ fixed to their SM values---the parametrization adopted throughout the main text for the interacting neutrino scenarios. For comparison, the dashed black curve shows the best-fit template when all three parameters are allowed to vary freely.

The good agreement between these two approaches, particularly at the higher multipoles most relevant for phase-shift measurements, demonstrates that the essential physics of neutrino interactions is captured by a simple amplitude rescaling of the SM template shape. This result validates our simplified parametrization and forms the basis for the systematic analysis presented in the next section, where we examine the regime of validity for this constant-amplitude approximation across different interaction models.

\subsection{Validation of the phase-shift template approximation}
\noindent Throughout the main text, we parametrize the phase shift induced by interacting neutrinos using a simple amplitude rescaling of the SM template $f_\ell^{\rm SM\nu}$; see Eq.~\eqref{eq:fell_int}. In this section, we examine the regime of validity for this approximation and quantify potential systematic effects. We first investigate how it performs for different power-law temperature dependencies of the interaction rate, then address the minor systematic effects from including temperature power spectrum in the template extraction.

\subsubsection{Limits of the constant amplitude approximation}

\begin{figure}
    \centering
    \includegraphics[width=\linewidth]{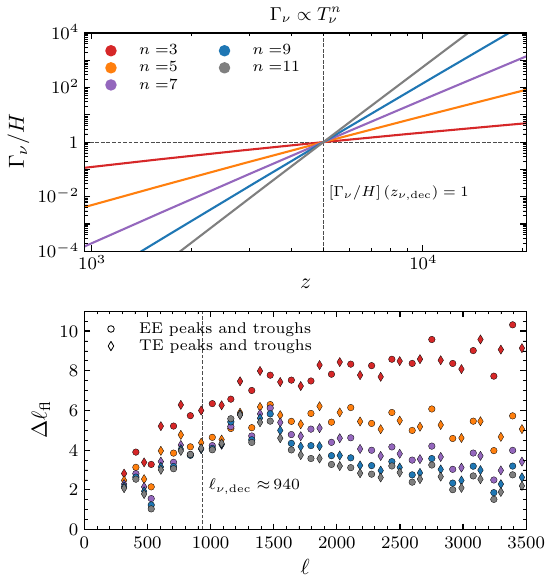}
    \caption{{\it Top:} Evolution of the ratio $\Gamma_\nu/H$ around the decoupling epoch for neutrino interaction scenarios with $\Gamma_\nu\propto T_\nu^n$ and $n\in\{3,5,7,9,11\}$, displayed in the corresponding colors shown in the legend, all calibrated to achieve decoupling at $z_{\nu,\rm dec}=5000$. {\it Bottom:}  Relative multipole shifts $\Delta\ell_{\rm fl}$ in the peaks and troughs of the EE (circles) and TE (diamonds) power spectra for three interacting neutrinos with respect to fluid-like neutrinos, for the same interaction models shown above. The phase shift remains effectively constant at high multipoles for $n\leq 5$, validating our approximation for all physically motivated interaction models, while deviations only appear for steeper temperature dependencies.}
    \label{fig:A02}
\end{figure}

\noindent The constant-amplitude approximation employed throughout our analysis relies on the gradual nature of neutrino decoupling in realistic interaction scenarios, characterized by scattering rates that follow power-law temperature dependencies, $\Gamma_\nu \propto T_\nu^n$. This gradual transition results in an extended loosely-coupled regime where the neutrino propagation speed takes intermediate values between fluid and free-streaming limits, producing a nearly constant phase shift whose amplitude reflects the ``average'' propagation speed during the relevant epochs. In the instantaneous decoupling limit, the phase shift drops abruptly from its asymptotic value to the fluid-like template for all multipoles $\ell > \ell_{\rm dec}$, with $\ell_{\rm dec}$ corresponding to modes entering the horizon at decoupling.

To quantify the regime of validity for our approximation, we have extended \texttt{nuCLASS} to accommodate arbitrary power-law temperature dependencies and computed the induced multipole shifts for varying $n$ at fixed decoupling redshift.  We find that our approximation remains robust for all $n \lesssim 5$ which encompasses the physically motivated neutrino interaction models of interest in this work. Deviations appear only for steeper temperature dependencies, which are difficult to construct theoretically, requiring additional model-building ingredients and symmetries~\cite{Boehm:2003hm,Stadler:2019dii}.

Figure~\ref{fig:A02} provides a representative illustration of our analysis for three neutrinos decoupling at $z_{\nu,\rm dec}=5000$ with $n \in \{3,5,7,9,11\}$. The bottom panel displays the relative shifts in the EE and TE power spectra with respect to fluid-like neutrinos. For $n \leq 5$, the phase shift stabilizes to an effectively constant value at high multipoles. The amplitude decreases with increasing $n$, since a higher temperature scaling results in stronger interactions prior to decoupling, making neutrinos increasingly more fluid-like during the epochs that imprint on CMB scales.  For $n > 5$, we observe a multipole dependence in the asymptotic behavior, with a gradual decrease in the phase shift amplitude for $\ell \gtrsim 1500$, becoming progressively more pronounced for larger $n$. Yet even in these extreme cases, the turnover occurs at significantly higher multipoles than the naive expectation of $\ell_{\rm dec} \approx 940$ from the instantaneous decoupling approximation, demonstrating the inherently extended nature of the decoupling transition in models with power-law temperature dependencies.

In the top panel of Fig.~\ref{fig:A02}, we show the evolution of $\Gamma_\nu/H$ over the redshift range $z \in [10^3, 2\times 10^4]$, around the decoupling epoch. For $n=3$, this ratio remains within one order of magnitude of unity throughout this range; for $n=5$, it spans at most two orders of magnitude. The gradual evolution is in sharp contrast to instantaneous decoupling, which forces interactions to cease abruptly at the decoupling redshift. Even for the extreme case of $n \geq 7$, where $\Gamma_\nu/H$ evolves over six orders of magnitude between $z\in[2\times 10^4,10^3]$, the ratio still remains within $10^{-2}-10^2$ for approximately 20\% of the redshift range around $z_{\nu,\rm dec}$. This remaining gradual component in the evolution of power-law interaction models explains why, even when our constant asymptotic-amplitude approximation begins to break down, the phase shift turnover remains smooth and displaced to higher multipoles compared to the instantaneous decoupling prediction.

\subsubsection{Impact of the temperature power spectra on phase shift extraction}
\begin{figure}
    \centering
    \includegraphics[width=\linewidth]{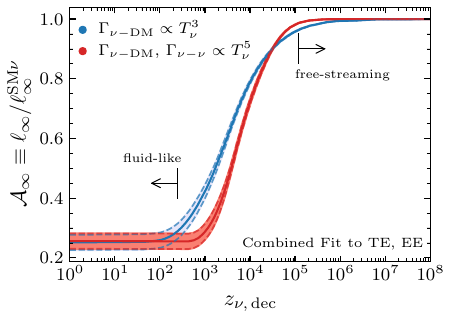}
    \caption{Same as Fig.~\ref{fig:3}, but using only the lensed TE and EE power spectra to extract the spectrum template fits $f_\ell$. In this case, both interaction models, with $\Gamma_\nu\propto T_\nu^3$ and $\Gamma_\nu\propto T_\nu^5$, converge to the same fluid-like limit at low $z_{\nu,\rm dec}$, confirming that the mild discrepancies seen in Fig.~\ref{fig:3} are numerical artifacts from imperfect phase shift isolation in the TT power spectrum.}
    \label{fig:A1}
\end{figure}

\noindent As noted in the main text, including the temperature power spectrum in our phase-shift template fits introduces minor systematic effects that warrant further discussion. While one would expect identical fluid-like limits regardless of the temperature scaling of the interaction rate, Fig.~\ref{fig:3} reveals small differences between the $\Gamma_\nu \propto T_\nu^3$ and $\Gamma_\nu \propto T_\nu^5$ scenarios at low $z_{\nu,\rm dec}$. These apparent discrepancies arise from the inherent complexity of the TT power spectrum's transfer function, which makes a clean isolation of the phase shift more challenging compared to the E-mode polarization and TE cross-correlation power spectra~\cite{Montefalcone:2025unv}.

Figure~\ref{fig:A1} demonstrates this effect explicitly by presenting the amplitude ratio $\mathcal{A}_\infty$ as a function of decoupling redshift $z_{\nu,\rm dec}$ when fitting only the EE and TE peaks and troughs. In this restricted analysis, the fluid-like limits for both interaction models converge to identical values, confirming that the differences observed in Fig.~\ref{fig:3} stem entirely from the inclusion of temperature power spectrum. The value of $\mathcal{A}_\infty$ in the fluid-like regime is slightly larger when using only EE and TE. As seen in Figs.~\ref{fig:1} and~\ref{fig:A0}, the TT power spectrum produces the largest outliers in the fluid-like regime, systematically overestimating the induced phase shift.

For self-consistency with the CMB data utilized in our observational analyses, we employ the full lensed-spectra fits throughout this work, including the TT contribution, despite these minor systematic effects. The resulting differences have negligible impact on our neutrino decoupling constraints, since the variations are well within the statistical uncertainties of the derived $\mathcal{A}_\infty$–$z_{\nu\rm dec}$ relationships. In addition, these differences manifest primarily at low decoupling redshifts close to the fluid-like limit, which we show to be strongly disfavored by current observations.

\section{Comparison of direct-phase-shift measurement methods}\label{sec:A3}
\begin{figure}
    \centering
    \includegraphics[width=\linewidth]{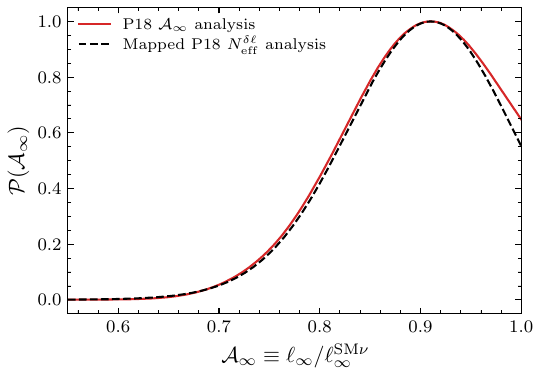}
    \caption{One-dimensional posterior of the phase-shift-amplitude ratio $\mathcal{A}_\infty$ from {\it Planck} 2018 (P18). The \textcolor{tabred}{red} solid line shows the constraints from this work, while the dashed black line shows those obtained by mapping the $N_{\rm eff}^{\delta\ell}$ posterior of Ref.~\cite{Montefalcone:2025unv} via Eq.~\eqref{eq:Ainfty_to_Ndphi}. The close agreement between the two demonstrates the robustness of phase shift measurements to different parameterizations and prior choices.}
    \label{fig:A2}
\end{figure}

\noindent In this appendix, we demonstrate the consistency between the phase-shift measurements obtained in this work, through the amplitude ratio $\mathcal{A}_\infty$, and those from Ref.~\cite{Montefalcone:2025unv}. The latter introduces an effective number of multipole-shifting relativistic species, $N_{\rm eff}^{\delta\ell}$, that exclusively controls the magnitude of the phase shift in CMB power spectra through an artificial multipole shift:
\begin{equation}
	\Delta\ell(N_{\rm eff}^{\delta\ell}, \Neff) = \bigl[A(N_{\rm eff}^{\delta\ell}) - A(\Neff)\bigr] f^{\rm SM\nu}_\ell\,,	\label{eq:deltaell_MWF}
\end{equation}
where $f_\ell^{\rm SM\nu}$ is the spectrum-based template from free-streaming neutrinos and $A(N_{\rm eff})$ is given by Eq.~\eqref{eq:ANeff}. Within this parametrization, assuming $N_{\rm eff}=N_{\rm eff}^{\rm SM}=3.044$, any evidence for deviations from SM expectations would manifest as a detection of $N_{\rm eff}^{\delta\ell} \neq N_{\rm eff}^{\rm SM}$ at high statistical significance. This method provides a consistency check on SM expectations; our analysis provides a direct constraint through the explicit $\mathcal{A}_\infty - z_{\nu,\rm dec}$ relationships derived in this work. Since both parameterizations affect only the amplitude of the induced phase shift, we can establish a one-to-one mapping between them. Combining Eqs.~\eqref{eq:deltaell_data} and~\eqref{eq:deltaell_MWF}, we obtain
\begin{equation}
    \mathcal{A}_\infty=\epsilon(N_{\rm eff}^{\delta\ell})/\epsilon(3.044), \label{eq:Ainfty_to_Ndphi}
\end{equation}
which reduces to $\mathcal{A}_\infty = 1$ for the SM prediction of $N_{\rm eff}^{\delta\ell} = 3.044$.

We apply this mapping to the posterior distribution of $N_{\rm eff}^{\delta\ell}$ from the {\it Planck} 2018-only analysis reported in Ref.~\cite{Montefalcone:2025unv} and compare it to our corresponding constraints on ${A}_\infty$ from the same dataset. Figure~\ref{fig:A2} displays the results, showing excellent agreement between the two approaches. Moreover, Ref.~\cite{Montefalcone:2025unv} employs a flat prior on $N_{\rm eff}^{\delta\ell} \in [0,6]$, which corresponds to a non-flat prior on $\mathcal{A}_\infty$, extending well above unity. In our work, we use a uniform prior on $\mathcal{A}_\infty \in [0,1]$. While the posteriors are nearly identical over most of the parameter space, minor differences do appear near $\mathcal{A}_\infty = 1$, where the mapped $N_{\rm eff}^{\delta\ell}$ posterior shows slightly less support, as expected from its broader prior range.

The demonstrated equivalence between our approach and that of Ref.~\cite{Montefalcone:2025unv} also opens an important avenue for future analyses. In addition to the spectrum-based method discussed above, Ref.~\cite{Montefalcone:2025unv} developed a complementary perturbation-based approach that directly measures the phase shift in the photon-baryon perturbations where it originates, using a wavenumber-dependent template parametrized by an effective number of phase-shifting species, $N_{\rm eff}^{\delta\phi}$. This more fundamental measurement was shown to have improved robustness and reduced sensitivity to systematic effects albeit maintaining consistency with the spectrum-based approach. While deriving a perturbation-based template directly for interacting neutrino scenarios would be technically challenging, one can leverage the established equivalence between $N_{\rm eff}^{\delta\phi}$ and $N_{\rm eff}^{\delta\ell}$ from Ref.~\cite{Montefalcone:2025unv}, combined with our mapping from $N_{\rm eff}^{\delta\ell}$ to $\mathcal{A}_\infty$, Eq.~\eqref{eq:Ainfty_to_Ndphi},
and the $\mathcal{A}_\infty - z_{\nu,\rm dec}$ relationships derived in this work (Fig.~\ref{fig:3}). This chain of mappings would allow one to constrain neutrino decoupling redshifts from the more robust perturbation-based analysis, providing an independent cross-check of our results with potentially improved systematic control.

\section{Additional data analyses}\label{sec:A4}

\noindent In this appendix, we present supplementary analyses that complement the main results of our paper. We first examine the consistency between the different {\it Planck} data releases by comparing our baseline {\it Planck} 2018 (P18) analysis with results obtained using the updated {\it Planck} 2021 (P21) PR4 likelihood. We then provide a comprehensive summary of phase shift constraints from all possible combinations of current CMB datasets, including individual analyses of SPT and ACT data as well as various joint analyses not featured in the main text.

\subsection{Comparison between {\it Planck} 2018 and {\it Planck} 2021 analyses}

\noindent In this section, we perform a comparison between the P18 analysis presented in the main text and one incorporating the latest P21 likelihood code based on the \texttt{NPIPE} data release~\cite{Planck:2020olo}. We replace the \texttt{plik-lite} high-$\ell$ likelihood with the updated \texttt{HiLLiPoP} likelihood~\cite{Tristram:2020wbi,Tristram:2023haj} and substitute the \texttt{SimAll} low-$\ell$ EE likelihood with the \texttt{LoLLiPoP} likelihood~\cite{Tristram:2020wbi,Tristram:2023haj}, while retaining the \texttt{Commander} likelihood for low-$\ell$ TT spectrum.

\begin{figure}
    \centering
    \includegraphics[width=\linewidth]{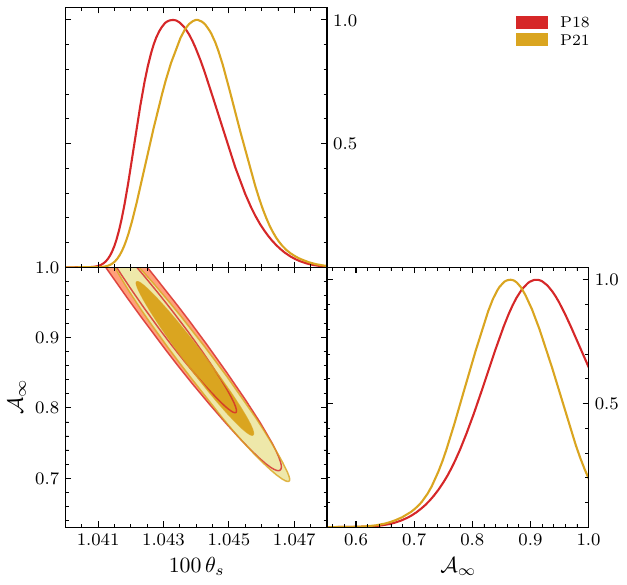}
    \caption{Comparison of constraints on the phase-shift-amplitude ratio $\mathcal{A}_\infty$ and the sound-horizon angle $\theta_s$ from {\it Planck} 2018 (P18, \textcolor{tabred}{red}) and {\it Planck} 2021 (P21, \textcolor{goldenrod}{gold}). Contours denote 68\% and 95\% confidence regions, with marginalized one-dimensional posteriors along the diagonal. The strong anti-correlation reflects their natural degeneracy, as $\theta_s$ and $\mathcal{A}_\infty$ shift the frequency and phase of acoustic oscillations, respectively. The constraints are broadly consistent between the two data releases, with only a minor shift toward smaller values of $\mathcal{A}_\infty$ in P21, worsening the consistency with the SM expectation ($\mathcal{A}_\infty = 1$) from $\sim 1.1\sigma$ to $\sim 1.8\sigma$.}
    \label{fig:A3}
\end{figure}

In Fig.~\ref{fig:A3}, we show the one-dimensional and two-dimensional posterior distributions for the phase-shift-amplitude ratio $\mathcal{A}_\infty$
and the angular size of the sound horizon $\theta_s$. We note that $\theta_s$ is the only $\Lambda$CDM parameter that exhibits significant degeneracy with $\mathcal{A}_\infty$. The strong anti-correlation between these two parameters is expected, since $\theta_s$ effectively measures the frequency of the acoustic oscillations.

The constraints on $\mathcal{A}_\infty$ are broadly consistent between the P18 and P21 analyses, though we observe a minor shift of the central value toward smaller amplitudes with P21. The P21 analysis yields $A_\infty=0.86_{-0.07}^{+0.08}$ and $A_\infty>0.74$ at the 68\%  and 95\% C.L. respectively, compared to $A_\infty=0.89_{-0.05}^{+0.10}$ and $A_\infty>0.76$ for P18. This shift, combined with the modest improvement in the $1\sigma$ confidence interval arising from reduced noise levels in P21, worsens the consistency with the SM expectation ($\mathcal{A}_\infty = 1$) from $\sim1.1\sigma$ in P18 to $\sim 1.8\,\sigma$ in P21. This minor discrepancy was previously identified in Ref.~\cite{Montefalcone:2025unv} in the context of $N_{\rm eff}^{\delta\ell}$ analyses and may be related to the well-documented $A_L$ anomaly~\cite{Planck:2018vyg}, which manifests as an elevated level of smoothing in the acoustic peaks and troughs of the {\it Planck} 2018 spectra compared to expectations based on the lensing power spectrum.

These differences have negligible impact on our combined dataset analyses. As noted in the main text, when incorporating ACT data, the multipole truncation at $\ell< 1000$ reduces sensitivity to the choice of {\it Planck} likelihood. This consideration is particularly relevant for our phase-shift analysis, which primarily relies on the positions of acoustic peaks and troughs at $\ell\gtrsim 1000$, where the phase shift signature is most pronounced. We, therefore, maintain our use of P18 in the combined analyses for consistency with the ACT collaboration's recommendations~\cite{ACT:2025fju}.

\subsection{Complementary dataset combinations}

\begin{table}[ht!]
	\centering
	\begin{tabular}{l c @{\hskip 2em} c @{\hskip 1em} c}
			\toprule
			& & \multicolumn{2}{c}{$z_{\nu,\rm dec}$}\\
            \cmidrule(lr){3-4}
	Dataset & {$\mathcal{A}_\infty$} & {$\Gamma_\nu\propto T_\nu^3$} & {$\Gamma_\nu\propto T_\nu^5$} \\
            \midrule[0.065em]
	SPT & $>0.75$	& $> 5.1\times 10^3$	& $>9.1\times 10^3$	\\[4pt]
        ACT & $>0.82$	& $> 7.9\times 10^3$	& $>1.27\times 10^4$	\\[4pt]
	P18 + SPT & $>0.83$	& $> 8.4\times 10^3$	& $>1.32\times 10^4$	\\[4pt]
        P18 + ACT & $>0.893$	& $> 1.30\times 10^4$	& $>1.69\times 10^4$ 	\\[4pt]
        P18 + ACT + SPT & $>0.897$	& $> 1.33\times 10^4$	& $>1.71\times 10^4$ \\[2pt]
	\bottomrule
	\end{tabular}
	\caption{Same as Table~\ref{tab:1} but for additional dataset combinations not presented in the main text: SPT only, ACT only, P18 + SPT, and P18 + ACT.  For completeness, we also include the P18 + ACT + SPT results, reporting $\mathcal{A}_\infty$ constraints to three decimal places for both P18 + ACT and P18 + ACT + SPT to highlight the marginal improvement from including SPT data in the combined analysis.}
	\label{tab:3}
\end{table}

\begin{figure}[hb!]
    \centering
    \includegraphics[width=\linewidth]{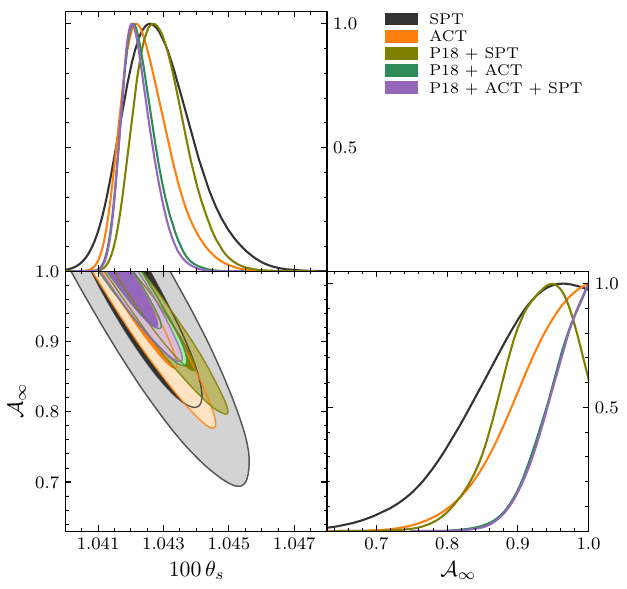}
    \caption{Constraints on the phase-shift-amplitude ratio $\mathcal{A}_\infty$ and the angular size of the sound horizon $\theta_s$ from various CMB datasets combinations, as indicated in the legend. The contours denote 68\% and 95\% confidence regions, with marginalized one-dimensional posteriors along the diagonal. All combinations show excellent agreement, yielding $\mathcal{A}_\infty$ consistent with the SM expectation ($\mathcal{A}_\infty=1$) at the $1\sigma$ level. Most notably, ground-based experiments provide constraints competitive with {\it Planck}, and ACT dominates the constraint due to its sensitivity to polarization power spectrum at the multipoles most relevant for phase shift measurements.}
    \label{fig:A4}
\end{figure}

\noindent In this section, we present a comprehensive analysis of phase-shift constraints from all possible combinations of {\it Planck} 2018, ACT, and SPT data beyond those featured in the main text. Figure~\ref{fig:A4} shows the posterior distributions for $\mathcal{A}_\infty$ and $\theta_s$ for these complementary dataset combinations, and Table~\ref{tab:3} summarizes the corresponding 95\% C.L. limits on $\mathcal{A}_\infty$, along with the derived constraints on the neutrino decoupling redshift for both the $\Gamma_\nu \propto T_\nu^3$ and $\Gamma_\nu \propto T_\nu^5$ universal interaction scenarios.

From Fig.~\ref{fig:A4}, we can clearly see the remarkable consistency between different CMB experiments, with each dataset combination yielding constraints on $\mathcal{A}_\infty$ that are consistent with the SM expectation at the $1\sigma$ level. In addition, all datasets exhibit the characteristic anti-correlation between $\mathcal{A}_\infty$ and $\theta_s$ discussed in the previous section.

Notably, both SPT and ACT provide competitive standalone constraints on the phase shift amplitude, highlighting the critical importance of high-precision polarization measurements for extracting this signature. ACT dominates the constraining power among ground-based experiments due to its exceptional sensitivity for the polarization power spectrum in the range $600\lesssim \ell \lesssim 2000$, which captures the multipoles most relevant for phase shift measurements.  This dominance is particularly apparent when comparing the P18 + ACT and P18 + ACT + SPT combinations; SPT yields negligible improvement to the constraints within our prior range.

Finally, as shown in Table~\ref{tab:3}, all dataset combinations consistently constrain universal neutrino interactions, requiring decoupling to occur deep within the radiation-dominated epoch, thus ruling out strongly interacting scenarios that would delay neutrino free-streaming into the matter-dominated era.

\newpage
\bibliographystyle{apsrev4-1}
\bibliography{bibl}
\end{document}